\documentstyle[twocolumn,epsf,aps,prb,floats]{revtex}
\begin{document}
\draft

\twocolumn[\hsize\textwidth\columnwidth\hsize\csname @twocolumnfalse\endcsname

\title{Current Characteristics of the Single-Electron Transistor
       at the Degeneracy Point}
\smallskip
\author{Amnon Buxboim and Avraham Schiller}
\address{Racah Institute of Physics, The Hebrew University,
	 Jerusalem 91904, Israel}
\date{\today}
\maketitle
\smallskip

\begin{abstract}
The linear and nonlinear transport properties of the single-electron
transistor at the degeneracy point are investigated for the case of
weak single-mode tunnel junctions. Two opposing scenarios are
considered,
distinguished by whether or not electrons can propagate coherently
between the two tunnel junctions. Each of these two scenarios
corresponds to the realization of a different multichannel Kondo
effect --- the two-channel Kondo effect in the case where coherent
propagation is allowed, and the four-channel Kondo effect in the
absence of coherent propagation. A detailed analysis of the linear
and nonlinear conductance is presented for each of these scenarios,
within a generalized noncrossing approximation for the nonequilibrium
multichannel Kondo Hamiltonian. A zero-bias anomaly is shown to
develop with decreasing temperature, characterized by the anomalous
power laws of the multichannel Kondo effect. A scaling function
of the differential conductance with $V/T$ ($V$ being the applied
voltage bias, $T$ the temperature) is proposed as a distinctive
experimental signature for each of these two scenarios. In the
absence of coherent propagation between the leads, and for
asymmetric couplings to the two leads, a crossover from four-channel
to two-channel behavior is manifested in a vanishing zero-temperature
conductance, and in a nonmonotonic voltage dependence of the
differential conductance for small asymmetries.
\end{abstract}

\smallskip
\smallskip

]
\narrowtext

\section{Introduction}
The single-electron transistor --- a small metallic or
semiconducting quantum box connected to two separate leads ---
is among the basic elements of mesoscopic devices. Due to
the finite energy barrier for charging the box with a single
electron, the so-called Coulomb blockade,~\cite{CB1,CB2}
charge inside the box is nearly quantized at low temperatures
for weak tunneling between the box and the leads. As a result,
transport through the box is strongly suppressed unless the
two lowest lying charge configurations in the box are tuned to
be degenerate. As a function of gate voltage, the conductance
thus shows a sequence of narrow peaks, each corresponding to
the crossing of the ground state from $n$ to $n+1$ excess
electrons inside the box.

At the degeneracy points, the system is subject to strong charge
fluctuations. The corresponding low-energy physics is governed
by the non-Fermi-liquid fixed point of the multichannel Kondo
effect,~\cite{Matveev91,Matveev95,CZ98} where the two degenerate
charge configurations in the box play the role of the impurity spin.
Which multichannel Kondo effect is realized depends on microscopic
details such as the number of transverse modes in the junctions,
and the nature of electron transport inside the box. Two opposing
scenarios were considered to date for the single-electron
transistor, distinguished by whether or not electrons can
propagate coherently between the two tunnel junctions. Focusing
on wide tunnel junctions, K\"onig {\em et al}.~\cite{KSS97}
considered the case where electrons can propagate coherently
between the two leads (see also Refs.~\onlinecite{SS94} and
\onlinecite{GKS+97}). Extending the work Grabert~\cite{Grabert94}
to nonequilibrium transport, these authors analyzed in detail
all second-order contributions to the current in the
dimensionless tunneling conductance, obtaining good agreement
with experiment.~\cite{Joyez_etal97} However, based on
perturbation theory, this approach breaks down near the
degeneracy points, where transport is governed at low
temperature by the strong electronic correlations of
the multichannel Kondo effect.

An alternative scenario was considered by Furusaki and
Matveev,~\cite{FM95} and later by Zar\'and {\em et al}.~\cite{ZZW00}
and Le Hur and Seelig.~\cite{LHS02}
Noting that elastic cotunneling is strongly suppressed at
temperatures above the level spacing,~\cite{AN90} these
authors omitted altogether coherent electron transport
between the leads, by coupling each lead to independent
conduction-electron modes within the box. For symmetric
single-mode junctions, the resulting low-temperature
physics is governed at the degeneracy point by the four-channel
Kondo effect,~\cite{FM95} in contrast to the two-channel
Kondo effect that takes place when electrons can propagate
coherently between the leads. Any
asymmetry in the coupling to the two leads drives the
system away from the four-channel fixed point to a
two-channel fixed point, where one lead is decoupled
from the box. Consequently, the zero-temperature conductance
vanishes, as shown by Furusaki and Matveev in the limit
of both a large asymmetry and strong tunneling to one
lead~\cite{FM95} (i.e., a nearly open tunneling mode).
A quantitative description of the temperature and
asymmetry dependence of the conductance in this case
remains lacking.

Despite considerable efforts,~\cite{KSS97,SS94,GKS+97,FM95,ZZW00,LHS02}
the understanding of the low-temperature transport at the
degeneracy points is far from complete for either scenario.
For weak single-mode tunnel junctions, there is no quantitative
theory for the temperature dependence of the conductance
in the Kondo regime, while the nonequilibrium differential
conductance is practically unexplored in this regime for
either scenario.~\cite{comment_on_diff_cond} The goal of
this paper is to provide a detailed analysis of the linear
and nonlinear transport at resonance, for weak single-mode
tunnel junctions. Both scenarios where electrons either can
or cannot propagate coherently between the two junctions are
considered. Our aim is to provide a host of signatures that
can be used to experimentally discern the two pictures, and
to detect which multichannel Kondo effect is realized in
actual systems.~\cite{BZAS99}

\begin{figure}[bt]
\centerline{\epsfxsize=60mm \epsfbox {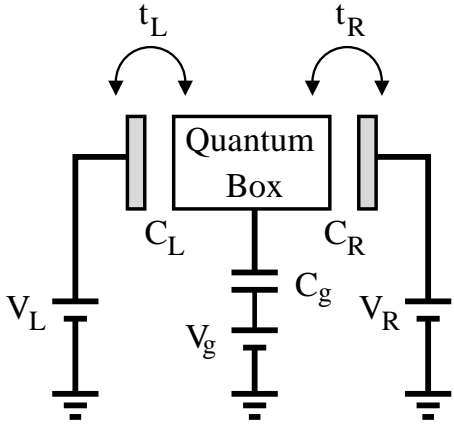}
}\vspace{8pt}
\caption{Schematic description of the physical system. A
         quantum box is coupled capacitively to a gate,
	 and connected by weak tunneling ($t_L$ and $t_R$)
	 to two metallic leads: a left lead and right lead.
	 The charge inside the box is controlled
	 by varying the gate voltage $V_g$, which fixes the
	 electrostatic potential within the box. A drain-source
	 voltage bias $V = V_R - V_L$ is applied across the
	 transistor, which drives a current through the device.}
\label{fig:fig1}
\end{figure}

To this end, we employ the noncrossing approximation~\cite{Bickers87}
(NCA). The NCA is a self-consistent perturbation theory
about the atomic limit. Originally designed to study dilute
magnetic alloys, this approach was successfully applied to
the out-of-equilibrium Kondo effect both for the
single-channel~\cite{WM94,HS95,NWML00}
and two-channel~\cite{HKH94} Anderson impurity model.
Recently, the NCA was generalized to the multichannel
Kondo spin Hamiltonian with arbitrary spin-exchange and
potential-scattering couplings.~\cite{LSZ01} Similar to the
NCA formulation of the multichannel Anderson model,~\cite{CR93}
the Kondo-NCA (KNCA) correctly describes the low-energy physics
of the multichannel Kondo model, reproducing the exact
non-Fermi-liquid power laws and logarithms of the multichannel
Kondo effect. This should be contrasted with the single-channel
case, where the NCA (and KNCA) fails to describe the Fermi-liquid
fixed point.~\cite{Bickers87} Here we extend the KNCA approach
to the nonequilibrium case, and apply it to the
single-electron transistor.

Using the KNCA, we find that a zero-bias anomaly develops
in the current characteristics of the single-electron
transistor, featuring the anomalous power laws of the
corresponding multichannel Kondo effect. The latter
power laws are manifest in the temperature and voltage
dependences of the low-temperature, low-bias
differential conductance. Similar to the case of
two-channel Kondo scattering off nonmagnetic two-level
tunneling systems,~\cite{HKH94,RLvDB94} these power laws
are best revealed in a scaling plot of the differential
conductance versus $e V/k_B T$, with $V$ being the applied
voltage bias. Such scaling plots are proposed as a distinct
experimental signature of the appropriate multichannel Kondo
effect. In the absence of electron propagation between the two
leads, a crossover from four-channel to two-channel behavior
is found in the differential conductance for asymmetric
couplings, in accordance with the picture of Furusaki and
Matveev.~\cite{FM95}

The remainder of the paper is organized as follows. In
sec.~\ref{sec:SET} we present the single-electron transistor,
and discuss the two alternative models under consideration. The
nonequilibrium KNCA is then formulated in sec.~\ref{sec:NCA},
followed by the derivation of the current in
sec.~\ref{sec:Current}. Our results for the linear and
nonlinear transport are presented in sec.~\ref{sec:Results},
and discuss in sec.~\ref{sec:Discussion}.

\section{The single-electron transistor}
\label{sec:SET}

\subsection{Basic model}

The physical system under consideration is shown schematically
in Fig.~\ref{fig:fig1}. A metallic island, or quantum box, is
connected by narrow point contacts to two separate leads, a
left lead ($L$) and a right lead ($R$). A drain-source voltage
bias is applied across the device, which sets a chemical-potential
difference between the leads: $\mu_L - \mu_R = e V$. Here $-e$
is the electron charge. Modeling each lead by $N$ noninteracting
one-dimensional conduction modes, the Hamiltonian of the two
independent leads reads
\begin{equation}
{\cal H}_{leads} = \sum_{\alpha = L, R} \sum_{n = 1}^{N}
	           \sum_{k, \sigma}
                        ( \epsilon_{\alpha k n} + \mu_{\alpha} )
	                c^{\dagger}_{\alpha k n \sigma}
			c_{\alpha k n \sigma} ,
\label{H_leads}
\end{equation}
where $c^{\dagger}_{\alpha k n \sigma}$ creates a conduction electron
with wave number $k$ and spin projection $\sigma$ in the $n$th mode
of lead $\alpha$, and $\epsilon_{\alpha k n}$ are the corresponding
single-particle energies, measured relative to the chemical potential
on that lead.

For the quantum box, one has to consider also its charging
energy. Setting the Fermi energy of the box as our reference
energy for the single-particle levels inside the box, the
excess number of electrons on the metallic island is described
by the operator
\begin{equation}
\hat{N}_{B} = \sum_{m = 1}^{M} \sum_{q, \sigma}
                \left[
		      c^{\dagger}_{B q m \sigma}c_{B q m \sigma} - 
		      \theta(-\epsilon_{B q m})
	      \right] .
\end {equation}
Here, similar to our notation for the leads,
$c^{\dagger}_{B q m \sigma}$ creates an electron with wave
number $q$ and spin projection $\sigma$ in the $m$th mode
of the box ($M$ independent conductance modes are taken
within the box), and $\epsilon_{B q m}$ are the corresponding
single-particle levels. The latter levels are assumed to
be sufficiently dense such that a continuum-limit description
can be used. The Hamiltonian of the isolated box is thus
given by
\begin{equation}
{\cal H}_{box} = \sum_{m = 1}^{M}
                 \sum_{q, \sigma} \epsilon_{B q m}
		      c^{\dagger}_{B q m \sigma}c_{B q m \sigma} 
               + E_C \left ( \hat{N}_{B} - N_C \right )^2 ,
\end{equation}
where $E_C$ is the charging energy of the box, and $N_C$ is the
classical number of excess electrons inside the box. These
two quantities are related to the capacitances of the gate
and of the left and right junctions through
$E_C = e^2/2(C_g + C_L + C_R)$ and
$N_C = (C_g V_g + C_L V_L + C_R V_R)/e$, where $V_g$, $V_L$,
and $V_R$ are the voltages applied to the gate and to the
left and right leads, respectively (see Fig.~\ref{fig:fig1}).

The quantum box is coupled to the leads by weak tunneling,
described by the tunneling Hamiltonian
\begin{equation}
{\cal H}_{tunnel} = \sum_{\alpha = L, R} \sum_{n, m}
                    \sum_{k, q, \sigma} t^{\alpha}_{n m}
		    \left\{
			 c^{\dagger}_{\alpha k n \sigma}
			 c_{B q m \sigma} + {\rm H.c.}
		    \right\} .
\label{H_tunnel}
\end{equation}
For simplicity, both the $k$ and $q$ dependences
of the tunneling matrix elements $t^L_{n m}$ and $t^R_{n m}$
have been neglected in Eq.~(\ref{H_tunnel}). The
full Hamiltonian of the system reads
${\cal H} = {\cal H}_{leads} + {\cal H}_{box} + {\cal H}_{tunnel}$.

\subsection{Mapping onto the multichannel Kondo problem}

Our objective is a detailed quantitative description of
the linear and nonlinear transport at resonance, when two
neighboring charge configurations are degenerate within the
box. To this end, let us focus on the vicinity of a particular
degeneracy point, $N_C = n + \frac{1}{2}$ with $n$ an integer,
separating the $\hat{N}_B = n$ and $\hat{N}_B = n+1$ charge
configurations.

As emphasized by Matveev,~\cite{Matveev91,Matveev95} the
low-temperature, low-bias physics is governed at resonance
by the intermediate-coupling fixed point of the multichannel
Kondo effect. To formulate the connection between the two
problems, we proceed along the lines laid out by
Matveev.~\cite{Matveev91} Labeling the deviation from the
degeneracy point by $\delta N = N_C - n - \frac{1}{2}$, we
concentrate on $|\delta N| \ll 1$ and $e |V|, k_B T \ll e^2/C_B$,
such that all charge configurations in the box other than
$\hat{N}_B = n$ and $\hat{N}_B = n+1$
are energetically inaccessible. One can formally remove
all higher energy charge configurations in the box by
means of two projection operators, $P_n$ and $P_{n+1}$,
which project onto the $\hat{N}_B = n$ and $\hat{N}_B = n+1$
subspaces, respectively. Alternatively, one can carry out
the projection onto the low-energy subspace by introducing
a spin-$\frac{1}{2}$ isospin operator $\vec{S}$, and identifying
\begin{eqnarray}
P_{n+1} - P_n &\longleftrightarrow& 2S_z ,
\label{S_z_def} \\
P_n c_{B q m \sigma} P_{n+1} &\longleftrightarrow&
                     c_{B q m \sigma}S^{-} ,\\
P_{n+1} c^{\dagger}_{B q m \sigma} P_n &\longleftrightarrow&
                     c^{\dagger}_{B q m \sigma}S^{+} .
\label{S^+_def}
\end{eqnarray}
Omitting an $E_C$-dependent reference energy, the resulting
low-energy Hamiltonian reads
\begin{eqnarray}
{\cal H}_{\rm eff} &=& \sum_{\alpha = L,R} \sum_{k, n, \sigma}
              (\epsilon_{\alpha k n} +\mu_{\alpha})
              c^{\dagger}_{\alpha k n \sigma} c_{\alpha k n \sigma}
\nonumber \\
&+& \sum_{q, m, \sigma} \epsilon_{B q m}
              c^{\dagger}_{B q m \sigma} c_{B q m \sigma}
	      - 2 E_C \delta N S_z
\nonumber \\
&+& \sum_{\alpha = L,R} \sum_{k, q, n, m, \sigma}
              \left\{ t^{\alpha}_{n m}
              c^{\dagger}_{\alpha k n \sigma} c_{B q m \sigma} S^{-}
              + {\rm H.c.} \right\} .
\label{H_eff}
\end{eqnarray}

Strictly speaking, Eqs.~(\ref{S_z_def})--(\ref{H_eff}) are
subject to the constraint $\hat{N}_B - S_z = n + 1/2$, reflecting
the fact that the isospin $\vec{S}$ and the conduction-electron
operators $c^{\dagger}_{B q m \sigma}$ are not independent degrees of
freedom. However, since the isospin dynamics in the Hamiltonian
of Eq.~(\ref{H_eff}) is not sensitive to the precise number of
conduction electrons inside the quantum box, this constraint can be
conveniently relaxed. We therefore regard hereafter the isospin
$\vec{S}$ and the $c^{\dagger}_{B q m \sigma}$ operators as
independent.

The Hamiltonian of Eq.~(\ref{H_eff}) has the form of a planner
multichannel Kondo Hamiltonian. Here $h = 2 E_C \delta N$
plays the role of a magnetic field, while the equivalence of
the two spin orientations guarantees the existence of at least
two identical conduction-electron channels. The total number
of independent conduction-electron channels depends, however,
on the microscopic details of the $t_{n m}^{\alpha}$ tunneling
matrix elements. In this paper we consider two particular
scenarios, corresponding to the two- and four-channel planner
Kondo Hamiltonian.

\subsection{Coherent propagation versus no coherent propagation
	    between the two junctions}

For sufficiently narrow point contacts, only a single
mode weakly couples each lead to the quantum box. Focusing
on this case, we consider two different scenarios: one by
which both leads couple to the {\em same} single mode within
the box, and the other whereby each lead is coupled to
a {\em different} mode within the box.

The former scenario is just a single-mode version of the model
of K\"onig {\em et al}.,~\cite{KSS97} whereby electrons can
propagate coherently between the two leads. Its corresponding
low-energy effective Hamiltonian is obtained by setting
$N = M = 1$ in Eqs.~(\ref{H_leads})--(\ref{H_tunnel}), thus
omitting the mode indices $n$ and $m$. The resulting
Hamiltonian takes the form
\begin{eqnarray}
{\cal H}_{\rm eff} &=& \sum_{\alpha = L,R} \sum_{k, \sigma}
     (\epsilon_{\alpha k} +\mu_{\alpha})
     c^{\dagger}_{\alpha k\sigma} c_{\alpha k\sigma} +
\sum_{q, \sigma} \epsilon_{B q}
      c^{\dagger}_{B q \sigma} c_{B q \sigma}
\nonumber \\
&+& \sum_{\alpha = L,R} \sum_{k, q, \sigma} t_{\alpha} \left\{
      c^{\dagger}_{\alpha k \sigma} c_{B q \sigma} S^{-}
      + {\rm H.c.} \right\} - h S_z ,
\label{H_2ch}
\end{eqnarray}
where $t^{L}_{n m} \to t_L$ and $t^{R}_{n m} \to t_R$ (both
taken to be real) are the tunneling matrix elements for the
left and right junctions, respectively.

The second scenario is a straightforward adaptation to weak
single-mode tunnel junctions of the model introduced by
Furusaki and Matveev,~\cite{FM95} and later used by Zar\'and
{\em et al}.~\cite{ZZW00} and by Le Hur and Seelig.~\cite{LHS02}
In this model, electron propagation between the two leads
is excluded from the outset, accounting thereby
for the strong suppression of elastic cotunneling at temperatures
above the level spacing.~\cite{AN90} The corresponding
low-energy effective Hamiltonian is obtained from
Eqs.~(\ref{H_leads})--(\ref{H_tunnel}) by setting $N = 1$
and $M = 2$, and taking $t^{L}_{1 m} \to t_L \delta_{m, 1}$
and $t^{R}_{1 m} \to t_R \delta_{m, 2}$. Converting for
convenience from the channel labels $m = 1$ and $m = 2$
within the box to $\alpha = L$ and $\alpha = R$, respectively,
the resulting Hamiltonian is given by
\begin{eqnarray}
{\cal H}_{\rm eff} &=& \sum_{\alpha = L, R} \sum_{k, \sigma}
     (\epsilon_{\alpha k} +\mu_{\alpha})
     c^{\dagger}_{\alpha k \sigma} c_{\alpha k \sigma}
\nonumber \\
&+& \sum_{\alpha = L, R} \sum_{q, \sigma} \epsilon_{B q \alpha}
      c^{\dagger}_{B q \alpha \sigma} c_{B q \alpha \sigma}
      - h S_z
\nonumber \\
&+& \sum_{\alpha = L,R} \sum_{k, q, \sigma} t_{\alpha} \left \{
      c^{\dagger}_{\alpha k \sigma} c_{B q \alpha \sigma} S^{-}
      + {\rm H.c.} \right\} .
\label{H_4ch}
\end{eqnarray}
As in Eq.~(\ref{H_2ch}), the effective magnetic field $h$ is
equal to $2 E_C \delta N$.

\subsubsection{Coherent propagation between the leads}

The Hamiltonian of Eq.~(\ref{H_2ch}) is equivalent in equilibrium
to the planner two-channel Kondo Hamiltonian in an applied
magnetic field. This is best seen by first converting to a
constant-energy-shell representation of the conduction-electron
creation and annihilation operators in each of the two leads
and the quantum box, and then constructing generalized
``bonding'' and ``anti-bonding'' combinations of the two leads.
Upon doing so the ``anti-bonding'' band decouples from the
isospin $\vec{S}$, while the ``bonding'' band plays the role
of the single lead in Matveev's original mapping onto the
planner two-channel Kondo Hamiltonian.~\cite{Matveev91}
At the degeneracy point, the capacitance of the single-electron
transistor diverges logarithmically with decreasing temperature
according to $C \propto (1/T_K)\log(T_K/T)$, where $T_K$ is
the two-channel Kondo temperature:~\cite{comment_on_Tk}
\begin{equation}
k_B T_K = \left ( D \sqrt{ g_L + g_R } \right )
          \exp \left [
              -\frac{\pi}{4\sqrt{ g_L + g_R } }
          \right ] .
\label{T_K_2ch}
\end{equation}
Here $D \sim 2 E_C$ is the effective conduction-electron bandwidth,
\begin{equation}
g_{\alpha} = \rho_{\alpha}(0) \rho_B(0) t_{\alpha}^{2}
\label{g_alpha_2ch}
\end{equation}
($\alpha = L, R$) are the dimensionless tunneling conductances
for the left and right junctions, and
\begin{equation}
\rho_{\gamma}(\epsilon) = \sum_k
             \delta(\epsilon - \epsilon_{\gamma k})
\end{equation}
are the underlying density of states ($\gamma = L, R$, or
$B$). Throughout this paper we use the notation by which
the argument of $\rho_{\gamma}(\epsilon)$ is measured
relative to the corresponding chemical potential, i.e.,
$\mu_L$ and $\mu_R$ for $\gamma = L$ and $R$, respectively,
and $\mu_B = 0$ for $\gamma = B$.

Away from equilibrium, it is still possible to construct
``bonding'' and ``anti-bonding'' combinations of the left
and right leads, yet one can no longer dismiss the
``anti-bonding'' degrees of freedom as irrelevant. While
the latter degrees of freedom remain decoupled from the
isospin $\vec{S}$ on the level of the Hamiltonian, they do
couple to $\vec{S}$ through the effective density matrix,
which assigns different chemical potentials to the right and
left leads. Hence both the ``bonding'' and ``anti-bonding''
combinations must be retained when computing the current
for a finite bias.

\subsubsection{No coherent propagation between the leads}
\label{subsec:4ch}

Contrary to the Hamiltonian of Eq.~(\ref{H_2ch}), the Hamiltonian
of Eq.~(\ref{H_4ch}) corresponds in equilibrium to the planner
four-channel Kondo Hamiltonian with channel anisotropy. Specifically,
there are two distinct pairs of equivalent channels, one pair
for each tunnel junction. A channel-isotropic four-channel Kondo
Hamiltonian is recovered only for equal tunneling conductances
for the left and right junctions, $g_L = g_R$. The latter
conductances are given by
\begin{equation}
g_{\alpha} = \rho_{\alpha}(0) \rho_{B \alpha}(0) t_{\alpha}^{2}
\label{g_alpha_4ch}
\end{equation}
($\alpha = L, R$), which differ from the expressions of
Eq.~(\ref{g_alpha_2ch}) only in the separate density
of states for the left and right modes within the quantum box:
\begin{equation}
\rho_{B \alpha}(\epsilon) = \sum_k
             \delta(\epsilon - \epsilon_{B k \alpha}) .
\end{equation}

For $g_L = g_R = g$, the low-energy physics of the transistor
is governed by the non-Fermi-liquid fixed point of the
four-channel Kondo effect. The corresponding Kondo temperature
is given by~\cite{comment_on_4ch_Tk}
\begin{equation}
k_B T_K = (D g) \exp \left [
              -\frac{\pi}{4\sqrt{ g } }
          \right ] ,
\label{T_K_4ch}
\end{equation}
where $D \sim 2 E_C$. Note that the exponent of Eq.~(\ref{T_K_4ch})
differs by a factor of $\sqrt{2}$ from that of Eq.~(\ref{T_K_2ch})
under the same condition that $g_L = g_R = g$. Any asymmetry in
the left and right tunneling conductances, $g_L \neq g_R$, drives
the system to a two-channel fixed point with one of the leads
effectively decoupled from the box.

\section{Noncrossing approximation}
\label{sec:NCA}

To obtain a reliable quantitative theory for the low-temperature
transport at resonance for each of the models of Eqs.~(\ref{H_2ch})
and (\ref{H_4ch}), we resort to a recent adaptation of the
noncrossing approximation (NCA) to the Kondo spin Hamiltonian
with arbitrary spin-exchange and potential-scattering
couplings.~\cite{LSZ01} In this section, we formulate the
Kondo-NCA (KNCA) for each of the models of Eqs.~(\ref{H_2ch})
and (\ref{H_4ch}), generalizing the KNCA to nonequilibrium.

\subsection{Slave-fermion representation}

To handle the isospin ${\vec{S}}$, which we refer to hereafter
as the impurity spin, we employ Abrikosov's slave-fermion
representation.~\cite{Abrikosov65} In this representation, one
assigns a pseudo fermion to each impurity spin state according to
\begin{eqnarray}
f_{+}^{\dagger} \left | 0 \right \rangle
                & \longleftrightarrow & \left | S_z = +1/2 \right \rangle ,
\\
f_{-}^{\dagger} \left | 0 \right \rangle
                & \longleftrightarrow & | S_z = -1/2 \rangle .
\end{eqnarray}
This assignment corresponds to the replacement of the impurity spin
operator by the bilinear pseudo-fermion operator
\begin{equation}
\vec{S} \longleftrightarrow \frac{1}{2} \sum_{\gamma \delta}
        f^{\dagger}_{\gamma} \vec{\sigma}_{\gamma\delta}f_{\delta} ,
\end{equation}
where $\vec{\sigma}$ are the Pauli matrices. The physical subspace
corresponds to the constraint $\hat{N}_f = \sum_{\gamma}
f^{\dagger}_{\gamma}f_{\gamma} = 1$, which represents the fact
that we are working within an enlarged Hilbert space. This
constraint distinguishes the pseudo fermions from ordinary
fermions.

The advantage of the slave-fermion representation stems from the
ability to use standard diagrammatic many-body techniques to
calculate physical observables. The difficulty lies in implementing
the constraint, which necessitates the introduction of a fictitious
``chemical potential'' $\lambda$ for the pseudo fermions. The latter
is taken to minus infinity at the end of the calculation, as
described below. For concreteness, let us focus in the following
on the case where coherent propagation is allowed between the two
junctions, i.e., the Hamiltonian of Eq.~(\ref{H_2ch}). The necessary
modifications for the model of Eq.~(\ref{H_4ch}) are detailed in
Appendix~\ref{app:KNCA_4ch}.

In conventional perturbation theory for a nonequilibrium problem,
one starts with an unperturbed system in equilibrium. All processes
that drive the system out of equilibrium are then switched on
adiabatically at some initial time $t_0$. For the problem at hand,
one starts with two decoupled leads, each with its own chemical
potential. The unperturbed Hamiltonian ${\cal H}_0$ thus lacks
the exchange term, and is given by
\begin{eqnarray}
{\cal H}_0 &=& \sum_{\alpha = L, R} \sum_{k, \sigma}
     (\epsilon_{\alpha k} +\mu_{\alpha})
     c^{\dagger}_{\alpha k\sigma} c_{\alpha k\sigma}
\nonumber \\
&+&
     \sum_{q, \sigma} \epsilon_{B q}
	  c^{\dagger}_{B q \sigma} c_{B q \sigma} +
     \sum_{\gamma = \pm} (\epsilon_{\gamma}-\lambda)
          f^{\dagger}_{\gamma} f_{\gamma}  .
\label{H_0}
\end{eqnarray}
Here $\epsilon_{\gamma}$ is equal to $-\gamma \frac{1}{2} h$, while
$\lambda$ is a fictitious chemical potential for the pseudo particles.
Note that ${\cal H}_0$ is bilinear and diagonal in the single-particle
operators $c^{\dagger}_{\alpha k \sigma}$ and $f^{\dagger}_{\gamma}$,
which makes it a suitable starting point for diagrammatic calculations.
The initial density matrix of the system is also diagonal in the
above single-particle operators, and has the form
\begin{equation}
\hat{\rho}_0 = \frac{e^{ -\beta ( {\cal H}_0 -\mu_L N_L -\mu_R N_R ) } }
    {{\rm Tr} \left \{ e^{-\beta({\cal H}_0 -\mu_L N_L -\mu_R N_R )}
\right \} } .
\label{rho_0}
\end{equation}

Due to the chemical potential $\lambda$ that was added to
${\cal H}_0$, the statistical weight of the $\hat{N}_f = m$
subspace has an extra factor of $e^{\beta \lambda m}$ in
Eq.~(\ref{rho_0}). This allows one to project out the
$\hat{N}_f = 1$, physical subspace. Specifically, the average
of any physical observable $\hat{O}$ can be expressed as
\begin{equation}
\langle \hat{O} \rangle_{\rm phys} = \frac{1}{Z_{\rm imp}}
	\lim_{\lambda \rightarrow -\infty} e^{-\beta\lambda}
	\langle \hat{O} \hat{N}_f \rangle_{\lambda} ,
\label{O_ave} 
\end{equation}
where
\begin{equation}
Z_{\rm imp} = \lim_{\lambda \rightarrow -\infty} e^{-\beta\lambda}
      \langle {\hat N}_f \rangle_{\lambda} .
\label{Z_imp}
\end{equation}
Here subscripts $\lambda$ denote averages with respect to the
enlarged Hilbert space. Since the $\hat{N}_f = 0$ subspace does
not contribute to the averages of Eqs.~(\ref{O_ave}) and
(\ref{Z_imp}) (other than through the normalization of
$\hat{\rho}_0$), then the leading-order terms in $e^{\beta \lambda}$
come from the $\hat{N}_f = 1$, physical subspace. The latter
terms are the only ones to survive the $\lambda \to -\infty$
limit. Note that, in practice, one can drop
the $\hat{N}_f$ operator from the average of Eq.~(\ref{O_ave}) for
those physical operators $\hat{O}$ that give zero when acting on the
$\hat{N}_f = 0$ subspace, which greatly simplifies the calculations.
We also note that $Z_{\rm imp}$ corresponds in equilibrium to the
``impurity contribution'' to the partition function (see, e.g.,
Ref.~\onlinecite{Bickers87}).

At $t_0$, the tunneling terms are switched on adiabatically,
and the system evolves according to the full Hamiltonian:
${\cal H}_0 + {\cal H}_{tun}$ with
\begin{equation}
{\cal H}_{tun} = \sum_{\alpha = L, R} \sum_{k, q, \sigma} t_{\alpha}
               \left \{
	           c^{\dagger}_{\alpha k \sigma} c_{B q \sigma}
		   f^{\dagger}_{-}f_{+} + {\rm H.c.}
               \right \} .
\end{equation}
After all transients have decayed, a new nonequilibrium steady-state
is reached, characterized by time-independent averages of physical
observables such as the current operator. Within the enlarged Hilbert
space, the steady-state average of such an operator ${\hat O}$ at
time $t = 0$ is given by
\begin{equation}
\langle \hat{O} \rangle_{\lambda} = \lim_{t_0 \to -\infty}
              {\rm Tr} \left\{
	      \hat{\rho}_0 {\hat U}^{\dagger}(0, t_0)
	      {\hat O} {\hat U}(0, t_0) \right \} ,
\end{equation}
where ${\hat U}$ is the time-evolution operator corresponding to the
full Hamiltonian. Projection onto the physical subspace is carried
out according to Eq.~(\ref{O_ave}).

\subsection{Noncrossing approximation}

The key ingredients for the calculation of physical observables are
the pseudo-fermion Green functions. These include the retarded and
advanced Green functions,
\begin{eqnarray}
G^{r (\lambda)}_{\gamma}(t,t') &=& -i\theta (t - t') 
        \left < \left\{
	         f_{\gamma}(t), f^{\dagger}_{\gamma}(t')
        \right\} \right >_{\lambda} , \\
G^{a (\lambda)}_{\gamma}(t,t') &=& i\theta (t' - t) 
        \left < \left\{
	         f_{\gamma}(t), f^{\dagger}_{\gamma}(t')
        \right\} \right >_{\lambda} ,
\end{eqnarray}
along with the lesser and greater Green functions,
\begin{figure}
\centerline{\epsfxsize=80mm \epsfbox {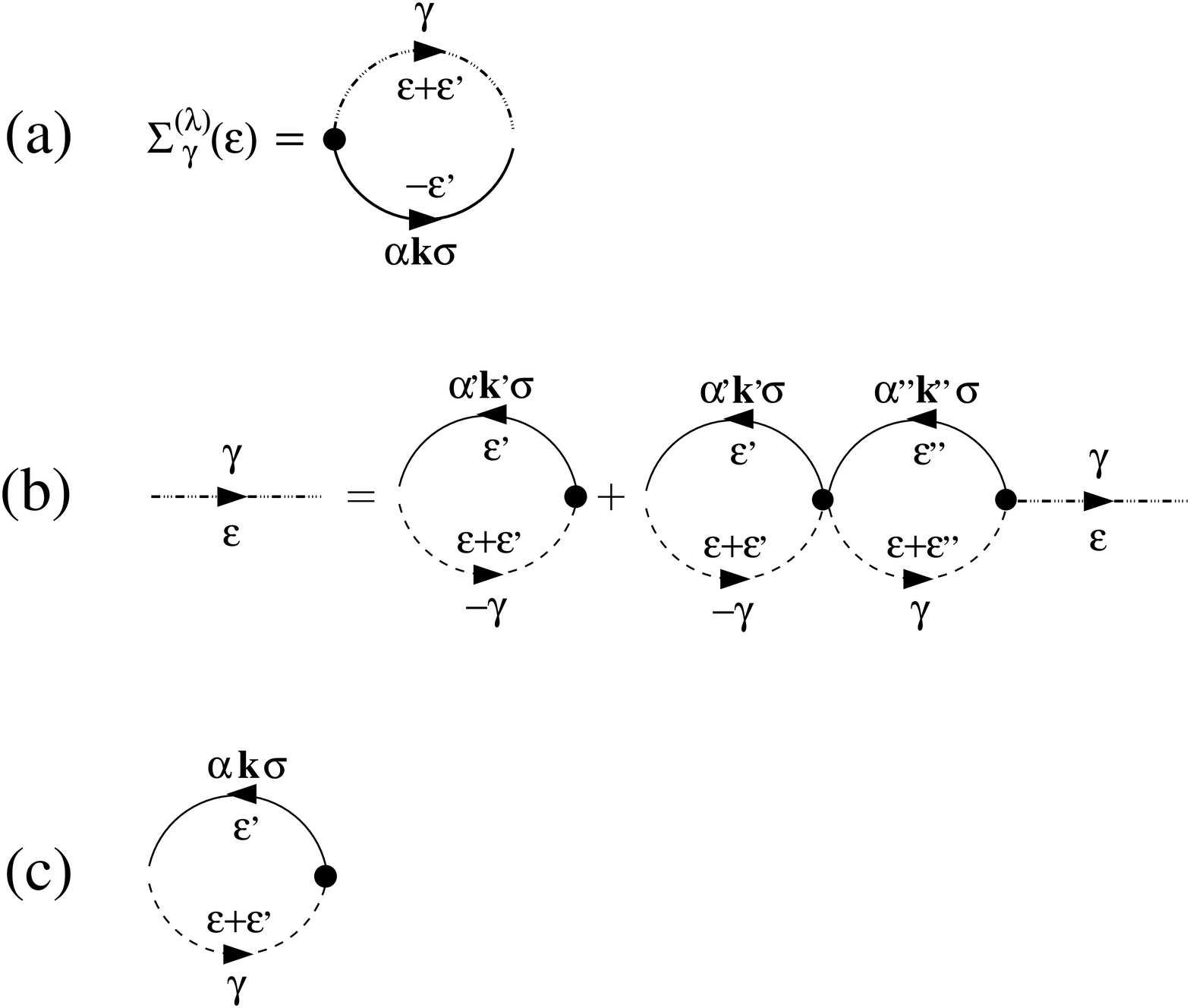}}
\vspace{10pt}
\caption{Diagrammatic representation of the KNCA approximation
         for the pseudo-fermion self-energies, for the
	 Hamiltonian of Eq.~(\ref{H_2ch}). Within the KNCA, the
	 pseudo-fermion self-energies, Fig.~(a), are approximated
	 by a bubble with one bare conduction-electron propagator
	 (full line) and one ladder propagator (dotted-dashed line).
	 The latter propagator is defined in Fig.~(b). Full
	 circles denote tunneling vertices. The ladder propagators
	 are constructed from particle-hole bubbles, Fig.~(c), with
	 one bare conduction-electron propagator and one fully
	 dressed pseudo-fermion propagator (dashed line). There
	 are two different bubbles:
	 $P^{(\lambda)}_-$, in which $\gamma = -$ and $\alpha = B$,
	 and $P^{(\lambda)}_+$, in which $\gamma = +$ and $\alpha$ is
	 summed over $L$ and $R$. The same convention for $\gamma$
	 and $\alpha$ applies to the self-energy bubble of Fig.~(a).
	 Note that we include the matrix elements of the two edge
	 vertices and a factor of $\rho_B(0)/g$ in our definition
	 of $P^{(\lambda)}_+$, while $P^{(\lambda)}_-$ is defined
	 without the matrix elements of the edge vertices, yet
	 with an extra factor of $1/\rho_B(0)$. Only particle-hole
	 bubbles are included within the KNCA, while all other
	 noncrossing diagrams (e.g., the particle-particle bubbles)
	 are omitted.}
\label{fig:fig2}
\end{figure}
\begin{eqnarray}
G^{< (\lambda)}_{\gamma}(t,t') &=& 
        \langle 
	         f^{\dagger}_{\gamma}(t') f_{\gamma}(t)
        \rangle_{\lambda} , \\
G^{> (\lambda)}_{\gamma}(t,t') &=& 
        \langle 
	         f_{\gamma}(t) f^{\dagger}_{\gamma}(t')
        \rangle_{\lambda} .
\end{eqnarray}
Here curly brackets denote the anticommutator, and $\gamma = \pm$.
Once steady state is reached, the above Green functions regain
time-translational invariance, and are solely dependent on the
time difference $\Delta t = t - t'$. It is therefore
advantageous to switch over to the energy domain, by introducing
the Fourier transforms with respect to $\Delta t/\hbar$. In
equilibrium, the lesser Green function is simply equal to the
spectral part of the retarded Green function times $2 \pi f(\epsilon)$,
where $f(\epsilon)$ is the Fermi-Dirac distribution function. Away from
equilibrium, when the effective distribution function is not known,
both the retarded and lesser Green functions are explicitly needed
in order to compute physical observables. 

In practice, the pseudo-fermion Green functions enter the calculation
of physical observables in their projected forms, which read
\begin{eqnarray}
G^{r}_{\gamma}(\epsilon) &=& \lim_{\lambda \rightarrow -\infty}
        G^{r (\lambda)}_{\gamma} (\epsilon-\lambda) ,
\label{Project_r}
\\
G^{a}_{\gamma}(\epsilon) &=& \lim_{\lambda \rightarrow -\infty}
        G^{a (\lambda)}_{\gamma} (\epsilon-\lambda) ,
\\
G^<_{\gamma} (\epsilon) &=& \lim_{\lambda \rightarrow -\infty} 
        e^{-\beta\lambda} G^{< (\lambda)}_{\gamma} (\epsilon-\lambda) ,
\label{G^<_projection}
\\
G^>_{\gamma} (\epsilon) &=& \lim_{\lambda \rightarrow -\infty}
        G^{> (\lambda)}_{\gamma} (\epsilon-\lambda) .
\label{Project_>}
\end{eqnarray}
These projections are analogous to the ones used in equilibrium,
when $G^{<}$ coincides with the negative-frequency, or ``defect'',
spectral function.~\cite{Bickers87} Note that, unlike $G^>$ and $G^r$,
the lesser Green function $G^{<}$ has no contribution from the
$\hat{N}_f = 0$ subspace. Rather, its leading-order term comes
from the $\hat{N}_f = 1$ subspace, which decays to zero as
$e^{\beta\lambda}$. The extra $e^{-\beta \lambda}$ exponent in
Eq.~(\ref{G^<_projection}) is responsible for canceling this
decay to zero when the limit $\lambda \rightarrow -\infty$ is
implemented.

The projected Green functions have standard forms in terms of the
projected self-energies. Specifically, $G^{r, a}_{\gamma}(\epsilon)$
and $G^{<, >}_{\gamma} (\epsilon)$ are equal to
\begin{eqnarray}
G^{r, a}_{\gamma} (\epsilon) &=&
        \frac{1}{\epsilon - \epsilon_{\gamma} -
	\Sigma_{\gamma}^{r, a}(\epsilon)} ,
\\
G^{<, >}_{\gamma} (\epsilon) &=&
        \Sigma^{<, >}_{\gamma}(\epsilon) 
        \left|  G^{r}_{\gamma} (\epsilon) \right|^2 ,
\end{eqnarray}
where $\Sigma_{\gamma}^{r, a}(\epsilon)$ and
$\Sigma^{<, >}_{\gamma}(\epsilon)$ are obtained from their
unprojected counterparts according to the projection rules
of Eqs.~(\ref{Project_r})--(\ref{Project_>}). The
KNCA~\cite{LSZ01} consists of a particular set of diagrams
for the pseudo-fermion self-energies, depicted in
Fig.~\ref{fig:fig2}. Figure~\ref{fig:fig2}(c) shows the building
block for the NCA self-energy diagrams of Fig.~\ref{fig:fig2}(a).
It consists of a particle-hole bubble, with one fully dressed
pseudo-fermion line and one bare conduction-electron line. The
KNCA self-energy features a ladder of such bubble diagrams,
Fig.~\ref{fig:fig2}(b), with vertices inserted in between.
Each ladder has an odd number
of bubbles, as the incoming and outgoing pseudo-fermion lines
share the same isospin label. For the Hamiltonian of
Eq.~(\ref{H_2ch}), the physical spin is conserved along the
ladder, whereas both the physical spin and the lead index
(i.e., $\alpha = L, R$) are conserved for the Hamiltonian of
Eq.~(\ref{H_4ch}).

Focusing on the Hamiltonian of Eq.~(\ref{H_2ch}), there are
two separate ladders, $D^{(\lambda)}_{\pm}$, labeled by the
index of the outgoing and incoming pseudo-fermion lines.
Projecting $D^{(\lambda)}_{\pm}$ according to the rules laid
out in Eqs.~(\ref{Project_r})--(\ref{Project_>}), one obtains
\begin{equation}
D^{r}_{\pm}(\epsilon) =
           \frac{g P^{r}_{\mp}(\epsilon)}
           {1 - g P^{r}_+(\epsilon)P^{r}_-(\epsilon)} ,
\label{D^r_2ch}
\end{equation}
\begin{equation}
D^{<}_{\pm}(\epsilon) =
           \frac{g P^{<}_{\mp}(\epsilon) + g^2
		  P^{<}_{\pm}(\epsilon)
		  \left| P^r_{\mp}(\epsilon) \right|^2}
           {\left| 1 - g P^{r}_+(\epsilon)
		  P^{r}_-(\epsilon)\right|^2} ,
\label{D^<_2ch}
\end{equation}
where $g_L$ and $g_R$ are the dimensionless tunneling
conductances for the left and right junction, defined in
Eq.~(\ref{g_alpha_2ch}), $g = g_L + g_R$ is the sum of
the two dimensionless conductances, and
\begin{eqnarray}
P^r_+(\epsilon) &=& - \!\!\sum_{\alpha = L, R} \frac{g_{\alpha}}{g}
      \int_{-\infty}^{\infty} G^r_+(\epsilon')
      f(\epsilon'\!-\!\epsilon\!-\!\mu_{\alpha})
\nonumber \\
&& \;\;\;\;\;\;\;\;\;\;\;\;\;\;\;\;\;\;\;\;\;\;\;\;\;
      \times\, \nu_{\alpha}(\epsilon'\!-\!\epsilon\!-\!\mu_{\alpha})
      d\epsilon' ,
\\
P^<_+(\epsilon) &=& - \!\!\sum_{\alpha = L, R} \frac{g_{\alpha}}{g}
      \int_{-\infty}^{\infty} G^<_+(\epsilon') 
      f(-\epsilon'\!+\!\epsilon\!+\!\mu_{\alpha})
\nonumber \\
&& \;\;\;\;\;\;\;\;\;\;\;\;\;\;\;\;\;\;\;\;\;\;\;\;\;
      \times\, \nu_{\alpha}(\epsilon'\!-\!\epsilon\!-\!\mu_{\alpha})
      d\epsilon' ,
\\
P^r_-(\epsilon) &=& -\!\int_{-\infty}^{\infty}
      G^r_-(\epsilon') f(\epsilon'\!-\!\epsilon)
      \nu_B(\epsilon'\!-\!\epsilon) d\epsilon' ,
\\
P^<_-(\epsilon) &=& -\!\int_{-\infty}^{\infty}
      G^<_-(\epsilon') f(-\epsilon'\!+\!\epsilon)
      \nu_B(\epsilon'\!-\!\epsilon) d\epsilon' .
\label{P^<_2ch_}
\end{eqnarray}
Here $\nu_{\gamma}(\epsilon) =
\rho_{\gamma}(\epsilon)/\rho_{\gamma}(0)$ with $\gamma = L, R$,
or $B$ is the reduced density of states.

In Eqs.~(\ref{D^r_2ch}) and (\ref{D^<_2ch}) we adopted the
convention by which the $D_{-}$ ladder includes the matrix
elements of the two end-point vertices, and is multiplied
by $\rho_B(0)$. The $D_{+}$ ladder does not include the
matrix elements of the two end-point vertices, but is
multiplied by $g/\rho_B(0)$. We further restricted
attention to the case of a zero physical magnetic field (not
to be confused with $h = 2 E_C \delta N$), in which case all
dependences on the physical spin index $\sigma$ drop. With
these notations the KNCA self-energies take the form
\begin{eqnarray}
\Sigma^r_+(\epsilon) &=& -2 \! \sum_{\alpha = L, R}
       \frac{g_{\alpha}}{g}
       \int_{-\infty}^{\infty} D^r_+(\epsilon') 
       f(-\epsilon\!+\!\epsilon'\!+\!\mu_{\alpha})
\nonumber \\
&& \;\;\;\;\;\;\;\;\;\;\;\;\;\;\;\;\;\;\;\;\;\;\;\;\;
       \times \,
       \nu_{\alpha}(\epsilon\!-\!\epsilon'\!-\!\mu_{\alpha})
       d\epsilon' ,
\\
\Sigma^<_+(\epsilon) &=& -2 \! \sum_{\alpha = L, R}
       \frac{g_{\alpha}}{g}
       \int_{-\infty}^{\infty} D^<_+(\epsilon') 
       f(\epsilon\!-\!\epsilon'\!-\!\mu_{\alpha})
\nonumber \\
&& \;\;\;\;\;\;\;\;\;\;\;\;\;\;\;\;\;\;\;\;\;\;\;\;\;
       \times \,
       \nu_{\alpha}(\epsilon\!-\!\epsilon'\!-\!\mu_{\alpha})
       d\epsilon' ,
\\
\Sigma^r_-(\epsilon) &=& -2 \! \int_{-\infty}^{\infty}
       D^r_-(\epsilon')f(-\epsilon\!+\!\epsilon')
       \nu_B(\epsilon\!-\!\epsilon') d\epsilon' ,
\\
\Sigma^<_-(\epsilon) &=& -2 \! \int_{-\infty}^{\infty}
       D^<_-(\epsilon')f(\epsilon\!-\!\epsilon')
       \nu_B(\epsilon\!-\!\epsilon') d\epsilon' ,
\label{S^<_2ch}
\end{eqnarray}
where an extra factor of two comes from summation over the
two equivalent spin orientations.

Equations~(\ref{D^r_2ch})--(\ref{S^<_2ch}) represent the
complete summation of the KNCA class of diagrams for the
pseudo-fermion self-energies, in the case where coherent
propagation is allowed between the two leads [i.e., the
Hamiltonian of Eq.~(\ref{H_2ch})]. For $g \nu_B(\epsilon)
= g_L \nu_L(\epsilon) + g_R \nu_R(\epsilon)$,~\cite{comment_on_nu's}
these equations properly reduce in equilibrium to those
of Ref.~\onlinecite{LSZ01} for the planner two-channel
Kondo Hamiltonian.~\cite{comment_on_correspondence}
The adaptation of Eqs.~(\ref{D^r_2ch})--(\ref{S^<_2ch})
to the Hamiltonian of Eq.~(\ref{H_4ch}) is specified in
Appendix~\ref{app:KNCA_4ch}.

\section{Formulation of the Current}
\label{sec:Current}

Our next goal is to formulate the current in terms of the
pseudo-fermion Green functions introduced in the previous
section. We begin our discussion with the model of
Eq.~(\ref{H_2ch}).

\subsection{Coherent propagation between the two junctions}

The operator $\hat{I}_{\alpha}$, describing the electrical current
flowing into lead $\alpha$, is given by the time derivative of
the charge operator for that lead, $\hat{Q}_{\alpha}$. Within
the slave-fermion representation of the Hamiltonian of
Eq.~(\ref{H_2ch}) one obtains
\begin{equation}
\hat{I}_{\alpha} = i \frac{e}{\hbar} t_{\alpha}
	   \sum_{k, k', \sigma} \left \{
	    c^{\dagger}_{\alpha k \sigma} c_{B k' \sigma}
	    f^{\dagger}_{-}f_{+} - {\rm H.c.} \right \} .
\label{I_operator}
\end{equation}
Applying the projection procedure of Eq.~(\ref{O_ave}), the
time-averaged current $I_{\alpha}(V)$ reads
\begin{equation}
I_{\alpha}(V) = -2 \frac{e}{\hbar} t_{\alpha} \sum_{\sigma}
         {\rm Im} \left \{ {\cal G}^{<}_{\alpha \sigma}(t,t) \right \} ,
\label{I_no_1}
\end{equation}
where ${\cal G}^{<}_{\alpha \sigma}(t,t')$ is equal to
\begin{equation}
{\cal G}^{<}_{\alpha \sigma}(t,t') = \frac{1}{Z_{\rm imp}}
         \lim_{\lambda \to -\infty} e^{-\beta \lambda}
         \sum_{k, k'} \left \langle 
            c^{\dagger}_{\alpha k' \sigma}(t')
	    \hat{F}_{k \sigma}(t) \right \rangle_{\lambda}
\label{cal_G}
\end{equation}
with
\begin{equation}
\hat{F}_{k \sigma}(t) = c_{B k \sigma}(t)
	                 f^{\dagger}_{-}(t) f_{+}(t) .
\label{F_ks}
\end{equation}

In steady state, ${\cal G}^{<}_{\alpha \sigma}(t,t')$ depends
solely on the time difference $\Delta t = t - t'$. Switching
over to the energy domain and applying standard perturbation
theory one obtains the exact relation
\begin{equation}
{\cal G}^{<}_{\alpha \sigma}(\epsilon) = t_{\alpha}
        \left [
            G^r_{\rm imp}(\epsilon)
	    \Gamma_{\alpha}^{<}(\epsilon\!-\!\mu_{\alpha}) +
            G^<_{\rm imp}(\epsilon)
            \Gamma_{\alpha}^{a}(\epsilon\!-\!\mu_{\alpha})
        \right ] ,
\label{introducing_G_imp}
\end{equation}
where $G_{\rm imp}^{<}(\epsilon)$ is the Fourier transform of
the response function
\begin{equation}
G_{\rm imp}^{<}(t, t') = \frac{1}{Z_{\rm imp}}
        \lim_{\lambda \to -\infty} e^{-\beta \lambda}
        \sum_{k, k'} \left  \langle
	         \hat{F}^{\dagger}_{k' \sigma}(t')
		 \hat{F}_{k \sigma}(t) \right \rangle_{\lambda} ,
\label{G_imp_t}
\end{equation}
$G_{\rm imp}^{r}(\epsilon)$ is its retarded counterpart, and
$\Gamma^<_{\alpha}(\epsilon)$ and $\Gamma^r_{\alpha}(\epsilon)$
are equal to
\begin{equation}
\Gamma^<_{\alpha}(\epsilon) =
        2\pi \rho_{\alpha}(\epsilon) f(\epsilon) ,
\end{equation}
\begin{equation}
\Gamma^a_{\alpha}(\epsilon) = \int_{-\infty}^{\infty}
        \frac{\rho_{\alpha}(\epsilon')}{\epsilon\!-\!\epsilon'\!-\!i\delta}
	   d\epsilon' .
\end{equation}

Physically, $T_{\alpha\alpha}(\epsilon) =
t_{\alpha}^2 G_{\rm imp}^{r}(\epsilon)$ is the conduction-electron
$T$-matrix for scattering from lead $\alpha$ to lead $\alpha$.
Hence $G_{\rm imp}$ is analogous to the dressed impurity Green
function in the Anderson impurity model. Indeed, inserting
Eq.~(\ref{introducing_G_imp}) into Eq.~(\ref{I_no_1}) yields
\begin{eqnarray}
I_{\alpha}(V) = \frac{2 e g_{\alpha}}{\hbar \rho_B(0)}
              \int_{-\infty}^{\infty} &&
	      \left [
	            2 \pi A_{\rm imp}(\epsilon)
		    f(\epsilon - \mu_{\alpha}) -
		    G^<_{\rm imp}(\epsilon)
	      \right ]
\nonumber \\
           && \times \, \nu_{\alpha}(\epsilon - \mu_{\alpha}) d\epsilon ,
\label{I_2ch_exact}
\end{eqnarray}
which has the same structure as the expression for the current
through an Anderson impurity.~\cite{MW92} Here $A_{\rm imp}(\epsilon)
= - (1/\pi) {\rm Im} \{G_{\rm imp}^r (\epsilon) \}$ is the
spectral part of $G_{\rm imp}^r$.

\begin{figure}[bt]
\centerline{\epsfxsize=45mm \epsfbox {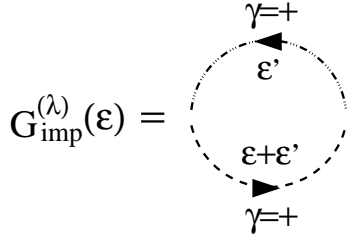}}
\vspace{10pt}
\caption{The KNCA diagram for $G_{\rm imp}$. Within the KNCA,
         $G_{\rm imp}$ reduces to a simple bubble, with one fully
	 dressed $f_{+}$ pseudo-fermion propagator (dashed line),
	 and one $D_{+}$ ladder propagator (dotted-dashed line).}
\label{fig:fig3}
\end{figure}

The energy integration in Eq.~(\ref{I_2ch_exact}) is restricted in
practice to an interval corresponding to the chemical-potential
difference between the leads, broadened by the temperature. To
the extent that one can neglect the energy dependence of
$\rho_{\alpha}(\epsilon)$ on that scale, it is possible to
eliminate $G_{\rm imp}^{<}$ from the expression for the current
by using the identity $I_L = -I_R  = (g_R I_L - g_L I_R)/g$, to
obtain
\begin{equation}
I_{L}(V) = \frac{2 e}{h}
              \int_{-\infty}^{\infty} 
              {\cal T}_{\rm eff}(\epsilon)
	      \left [
		    f(\epsilon - \mu_L) -
		    f(\epsilon - \mu_R)
	      \right ]
	      d\epsilon
\label{I_2ch}
\end{equation}
with
\begin{equation}
{\cal T}_{\rm eff}(\epsilon) = 
              \frac{4 \pi^2 g_L g_R}{\rho_B(0) g}
              A_{\rm imp}(\epsilon) .
\label{T_eff_2ch}
\end{equation}

Equation~(\ref{I_2ch}) is exact in the wide-band limit, when the
energy dependence of $\rho_{\alpha}(\epsilon)$ can be neglected.
It has the familiar form of an integral of an effective
transmission coefficient ${\cal T}_{\rm eff}(\epsilon)$ times
the difference of two Fermi functions. Similar to the case
of tunneling through an Anderson impurity,
${\cal T}_{\rm eff}(\epsilon)$ depends strongly on the
temperature and bias at low energies. Our main task is to
evaluate this effective transmission coefficient, which
requires knowledge of $G^r_{\rm imp}(\epsilon)$ and
$G^<_{\rm imp}(\epsilon)$. We calculate the latter functions
using the NCA class of diagrams.

Figure~\ref{fig:fig3} shows the KNCA diagram for $G_{\rm imp}$.
Similar to the standard NCA formulation of the impurity Green
function for the Anderson impurity model,~\cite{Bickers87}
$G_{\rm imp}$ is given by a simple bubble diagram, which consists
of one fully dressed $f_{+}$ pseudo-fermion propagator and one
$D_{+}$ ladder propagator. The resulting KNCA expressions for
$G^r_{\rm imp}(\epsilon)$ and $G^<_{\rm imp}(\epsilon)$ read
\begin{eqnarray}
G^r_{\rm imp}(\epsilon) &=& \frac{\rho_B(0)}{ 2 \pi g Z_{\rm imp} }
             \int_{-\infty}^{\infty} \left [
             G^<_+(\epsilon') D^a_+(\epsilon' - \epsilon) \right.
\nonumber \\
&& \;\;\;\;\;\;\;\;\;\;\;\;\;\;\;\;\;\;\;\;\;\;\; \left. -
             G^r_+(\epsilon + \epsilon') D^<_+(\epsilon') \right]
	     d\epsilon' ,
\label{G^r_imp_2ch}
\end{eqnarray}
\begin{equation}
G^<_{\rm imp}(\epsilon) = \frac{\rho_B(0)}{ \pi g Z_{\rm imp} }
             \int_{-\infty}^{\infty}
             G^<_+(\epsilon + \epsilon')
	     {\rm Im} \left\{ D^r_+(\epsilon') \right\}
	     d\epsilon' ,
\end{equation}
where
\begin{equation}
Z_{\rm imp} = \int_{-\infty}^{\infty}
              \left [
                      G_{+}^{<}(\epsilon) +
                      G_{-}^{<}(\epsilon)
              \right ]
	      \frac{d \epsilon}{2 \pi}\ .
\end{equation}
Hence $G^r_{\rm imp}(\epsilon)$ and $G^<_{\rm imp}(\epsilon)$
reduce within the KNCA to convolutions of the $\gamma = +$
pseudo-fermion Green function and the $D_{+}$ ladder.

\subsection{No coherent propagation between the two junctions}

Formulation of the current for the Hamiltonian of Eq.~(\ref{H_4ch})
follows the same basic steps presented in the previous subsection
for the Hamiltonian of Eq.~(\ref{H_2ch}).
The sole difference in Eqs.~(\ref{I_operator})--(\ref{F_ks}) enters
through the $c_{B k' \sigma}$ operators, which acquire an
additional lead index. The latter index carries over to
$\hat{F}_{k\sigma}$ and $G_{\rm imp}^{<}(t, t')$, whose definitions
are modified according to
\begin{equation}
\hat{F}_{\alpha k \sigma}(t) = c_{B k \alpha \sigma}(t)
	                 f^{\dagger}_{-}(t) f_{+}(t) ,
\end{equation}
\begin{equation}
G_{{\rm imp}, \alpha}^{<}(t, t') = \frac{1}{Z_{\rm imp}}
        \lim_{\lambda \to -\infty} e^{-\beta \lambda}
        \sum_{k, k'} \! \left \langle \!
	         \hat{F}^{\dagger}_{\alpha k' \sigma}(t')
		 \hat{F}_{\alpha k \sigma}(t)
	\right \rangle_{\lambda} \! .
\end{equation}
Contrary to Eq.~(\ref{I_2ch_exact}) for the Hamiltonian of
Eq.~(\ref{H_2ch}), each of the expressions for $I_L(V)$ and
$I_R(V)$ involve then a different $G_{{\rm imp}, \alpha}$
function:
\begin{eqnarray}
I_{\alpha}(V) = \frac{2 e g_{\alpha}}{\hbar \rho_{B \alpha}(0)}
              \int_{-\infty}^{\infty} \! &&
	      \left [
	            2 \pi A_{{\rm imp}, \alpha}(\epsilon)
		    f(\epsilon\!-\!\mu_{\alpha}) -
		    G^<_{{\rm imp}, \alpha}(\epsilon)
	      \right ]
\nonumber \\
           && \times \, \nu_{\alpha}(\epsilon - \mu_{\alpha})
	   d\epsilon .
\label{I_4ch_exact}
\end{eqnarray}
Here $g_{\alpha}$ are the dimensionless tunneling conductances
of Eq.~(\ref{g_alpha_4ch}), and $A_{{\rm imp}, \alpha}(\epsilon)$
is the spectral part of $G_{{\rm imp}, \alpha}^r$.

In general, $G_{{\rm imp}, L}$ and $G_{{\rm imp}, R}$ are two
distinct functions, with no simple relation. As a result,
one can no longer exploit current conservation to eliminate
$G^{<}_{{\rm imp}, \alpha}$ from the expression for the
current, as was done in Eq.~(\ref{I_2ch}). Hence, there
is no simple analog to the effective transmission
coefficient of Eq.~(\ref{T_eff_2ch}) when no coherent
propagation is allowed between the two leads. This is to be
expected, as electrons do not truly propagate between
the two leads in this case.

As before, $G_{{\rm imp}, L}(\epsilon)$ and
$G_{{\rm imp}, R}(\epsilon)$ are still given within the KNCA
by the bubble diagram of Fig.~\ref{fig:fig3}, which features,
however, a different ladder for $\alpha = L$ and $\alpha = R$.
Explicitly, one has
\begin{eqnarray}
G^r_{{\rm imp}, \alpha}(\epsilon) &=& \frac{\rho_{B \alpha}(0)}
             { 2 \pi g_{\alpha} Z_{\rm imp} }
             \int_{-\infty}^{\infty} \left [
             G^<_+(\epsilon') D^a_{\alpha +}(\epsilon' - \epsilon) \right.
\nonumber \\
&& \;\;\;\;\;\;\;\;\;\;\;\;\;\;\;\;\;\;\;\;\; \left. -
             G^r_+(\epsilon + \epsilon') D^<_{\alpha +}(\epsilon')
	     \right] d\epsilon' ,
\end{eqnarray}
\begin{equation}
G^<_{{\rm imp}, \alpha}(\epsilon) = \frac{\rho_{B \alpha}(0)}
             { \pi g_{\alpha} Z_{\rm imp} }
             \int_{-\infty}^{\infty}
             G^<_+(\epsilon + \epsilon')
	     {\rm Im} \left\{ D^r_{\alpha +}(\epsilon') \right\}
	     d\epsilon' .
\end{equation}

In the following section, we present our numerical results for the
differential conductance as obtained from Eqs.~(\ref{I_2ch_exact})
and (\ref{I_4ch_exact}) using the KNCA.

\section{Results}
\label{sec:Results}

We have computed the differential conductance for each of the
models of Eqs.~(\ref{H_2ch}) and (\ref{H_4ch}), by first solving
the KNCA equations for the pseudo-fermion Green functions and
ladders, and then evaluating the current using
Eqs.~(\ref{I_2ch_exact}) and (\ref{I_4ch_exact}), respectively.
The differential conductance $G(T, V) = dI/dV$ was obtained
by numerically differentiating the current with respect
to $V$, without resorting to the wide-band limit leading to
Eq.~(\ref{I_2ch}). The most difficult step in the above procedure
is solution of the KNCA equations, which requires simultaneous
solution of the retarded and lesser pseudo-fermion and ladder
functions. This was achieved by repeated iterations of the KNCA
equations until convergence is reached. As is always the case
with numerical solutions of NCA-type equations, the key to
high-precision numerics lies in a well-designed grid of mesh
points that scatter points more densely near the threshold and
peaks of the relevant functions. To this end, we have used a
combination of linear and logarithmic grids. As a critical test
for the precision of our numerical code, we have checked in all
our runs that the pseudo-fermion spectral functions fulfilled
the spectral sum rule to within one part in one thousand
($0.1$\%). Extreme precision was required at very low temperatures
and bias, when $k_B T$ and $|eV|$ were both much smaller
than the corresponding Kondo temperature $k_B T_K$.

Throughout this paper we focus our attention on the
degeneracy point, $h = 0$, and set
$\mu_L = -\mu_R = eV/2$.~\cite{comment_on_mu_ave}
We further use a semi-circular
conduction-electron density of states with half width $D$ for
the leads and the quantum box: $\rho_{\alpha}(\epsilon) =
\rho_{\gamma}(0) \sqrt{1 - (\epsilon/D)^2}$. Here $\gamma$ is
equal to $L, R$ or $B$ for the Hamiltonian of Eq.~(\ref{H_2ch}),
and $L, R, BL$ or $BR$ for the Hamiltonian of
Eq.~(\ref{H_4ch}). Note that we restrict ourselves for
simplicity to a single joint bandwidth $D$ for the box
and the leads. Although this need
not be the case in real systems, the precise details of the band
cut-offs should not affect the low-energy physics of interest
here.~\cite{Comment_on_D}

\subsection{Coherent propagation between the two junctions}

We begin our discussion with the case where coherent propagation
is allowed between the leads, i.e., the Hamiltonian of
Eq.~(\ref{H_2ch}). As discussed above, the latter Hamiltonian
reduces in equilibrium to the planner two-channel Kondo model
with $\rho_0 J_{\perp} = 2 \sqrt{g}$. Thus, the low-energy
physics of the model is governed by the corresponding Kondo
temperature $T_K$, which we extract by fitting the slope of
the $\ln(T)$ diverging term in the isospin susceptibility
\begin{equation}
\chi_{\rm isospin} = \left.
                     \frac{\partial \langle S_z \rangle}
                          {\partial h}
                     \right|_{h = 0}
\end{equation}
to the Bethe ansatz expression~\cite{SS89}
\begin{equation}
\chi_{\rm isospin}(T) \sim \frac{1}{20 k_B T_K}
                           \ln (T_K/T) .
\end{equation}
For $g = g_L + g_R = 0.04$, which corresponds to
$\rho_0 J_{\perp} = 0.4$ in the planner two-channel Kondo
representation, we obtain $k_B T_K/D = 3.85 \times 10^{-3}$.

As discussed at length in Ref.~\onlinecite{LSZ01}, the KNCA
estimate of $T_K$ deviates from the correct value of
Eq.~(\ref{T_K_2ch}). However, as shown by comparison with
the exact Bethe ansatz solution,~\cite{LSZ01,SS91} the
KNCA gives surprisingly accurate results for the temperature
and field dependence of the magnetic susceptibility in the
isotropic two-channel model, when $T$ and $\mu_B g_J H$ are
rescaled with the extracted Kondo temperature. We expect a similar
level of accuracy to hold for the nonequilibrium current in
each of the models under consideration.

\begin{figure} [tb]
\centerline{\epsfxsize=80mm \epsfbox{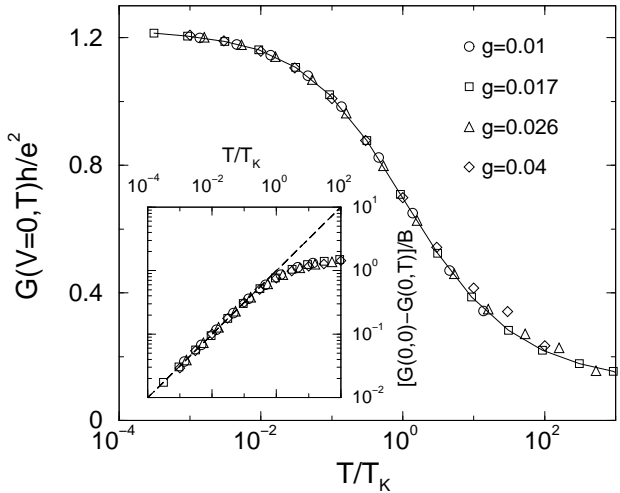}}
\vspace{5pt}
\caption{Temperature dependence of the zero-bias conductance,
         for the model of Eq.~(\ref{H_2ch}) with different
	 values of $g_L = g_R = g/2$. The corresponding Kondo
	 temperatures are equal to $k_B T_K/D =
	 8.4\times 10^{-6}, 1.25\times 10^{-4},
	 7.25\times 10^{-4}$, and $3.85\times 10^{-3}$ for
	 $g = 0.01, 0.017, 0.026$, and $0.04$, respectively.
	 With decreasing temperature, the conductance monotonically
	 increases according a universal curve (full line).
	 Below $T_K$, the conductance crosses over
	 to a $\sqrt{T}$ temperature dependence, in accordance with
	 the expected power-law behavior of the two-channel Kondo
	 effect. This is demonstrated in the inset, where a log-log
	 plot of $[G(0, 0) - G(0, T)]/B$ versus $T/T_K$ is shown
	 and compared to $\sqrt{T/T_K}$ (dashed line). Here
	 $G(0,0)$ and $B$ were obtained separately for each curve
	 from a fit to the $\sqrt{T}$ temperature dependence of
	 Eq.~(\ref{2ch_conductance}). All values of $G(0, 0)$
	 and $B$ so obtained fall within the range $G(0, 0) h/e^2
	 = 1.225 \pm 0.005$ and $Bh/e^2 = 0.68 \pm 0.015$. The
	 extracted values for $G(0, 0)h/e^2$ all fall within 1\%
	 from the exact zero-temperature KNCA value of $\pi^2/8$.}
\label{fig:fig4}
\end{figure}

\subsubsection{Symmetric coupling}

Figure~\ref{fig:fig4} shows the temperature dependence of the
zero-bias conductance for different values of $g_L = g_R$. As
expected of Kondo-assisted tunneling, the conductance is
enhanced with decreasing temperature according to a universal
curve. Although $T_K$ is varied by nearly three orders of
magnitude, all curves collapse onto a single line. Slight
deviations from universality are
seen above $T/T_K \approx 10$ for the larger couplings, which
may be due to the asymmetric line shape of the imaginary part
of the conduction-electron $T$-matrix within the KNCA (see,
e.g., Fig.~12 of Ref.~\onlinecite{LSZ01}). The latter asymmetry
is enhanced at high energies as the coupling is increased.

For over a decade of temperature around $T_K$, the enhancement
in $G(0, T)$ is approximately logarithmic in temperature. Below
$T_K$, the conductance crosses over to a $\sqrt{T}$ temperature
dependence. As demonstrated in the inset of Fig.~\ref{fig:fig4},
the low-temperature conductance can be successfully fit to the
form
\begin{equation}
G(0, T) = G(0, 0) - B \sqrt{ \frac{T}{T_K} }
\label{2ch_conductance}
\end{equation}
with $G(0,0)h/e^2 = 1.225 \pm 0.005$ and $Bh/e^2 = 0.68 \pm 0.015$.
Such a square-root temperature dependence of the conductance is
expected of the two-channel Kondo effect. It stems from the
square-root energy dependence of the imaginary part of the
zero-temperature, zero-bias conduction-electron
$T$-matrix,~\cite{CZ98,AL91} or $A_{\rm imp}(\epsilon)$ in this
context. A similar square-root temperature dependence of the
conductance was also found for tunneling through a two-channel Anderson
impurity,~\cite{HKH94} and measured for zero-bias anomalies in
ballistic metallic point contacts.~\cite{RB92} Note that the extracted
zero-temperature conductance corresponds to a mostly open channel
($61\%$ of the full conductance), and falls within 1\% from the
exact zero-temperature KNCA value of $G(0, 0)h/e^2 = \pi^2/8$.
The latter figure is obtained by taking the zero-temperature limit
of Eqs.~(\ref{D^r_2ch})--(\ref{S^<_2ch}) and (\ref{G^r_imp_2ch})
and performing a M\"uller-Hartmann type of analysis,~\cite{MH84}
reproducing the results of Cox and Ruckenstein for the NCA
treatment of the multichannel Anderson model.~\cite{CR93}

It should be emphasized that the above enhancement of the
zero-bias conductance for single-mode junctions differs
markedly from the case of wide tunnel junctions considered
by Schoeller and Sch\"on.~\cite{SS94} Omitting inter-mode
mixing, these authors found that $G(0, T)$ vanishes as
$1/\ln(T)$ at low temperatures, in accordance with the
vanishing-coupling fixed point of the infinite-channel
Kondo Hamiltonian. However, as recently shown by Zar\'and
{\em et al}.,~\cite{ZZW00} for realistic junctions with
inter-mode mixing there exists an exponentially small
crossover temperature $T^{\ast}$, below which two-channel
Kondo behavior prevails even for wide junctions. Hence
our results should also apply to wide tunnel junctions
at sufficiently low temperatures, provided electrons can
propagate coherently between the two leads.

\begin{figure}[tb]
\centerline{\epsfxsize=75mm \epsfbox {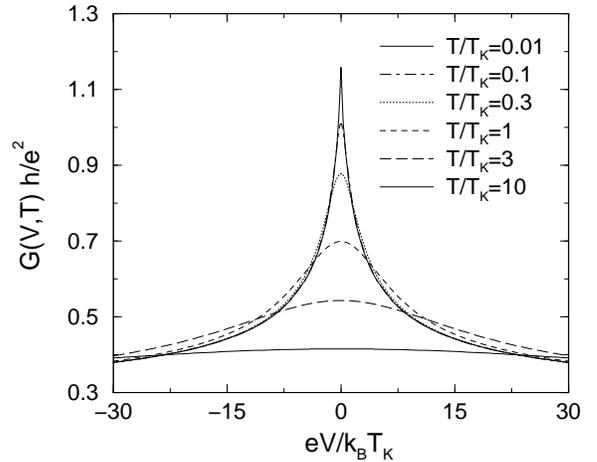}}
\vspace{5pt}
\caption{The differential conductance $G(V, T) = dI/dV$ versus
         voltage bias, for the model of Eq.~(\ref{H_2ch}) with
	 $g_L = g_R = 0.02$ and different temperatures.
	 The corresponding Kondo temperature is equal to
	 $k_B T_K/D = 3.85\times 10^{-3}$. With decreasing
	 temperature, a zero-bias anomaly develops in $G(V, T)$.
	 For $T < T_K$, the anomaly acquires a $\sqrt{V}$
	 voltage dependence, which is rounded off for voltages
	 on the scale of the temperature. For $T = 0$, a sharp
	 cusp is left in $G(V, 0)$ at zero bias.}
\label{fig:fig5}
\end{figure}

Figure~\ref{fig:fig5} shows the differential conductance
for the symmetric coupling $g_L = g_R = 0.02$ and different
temperatures. With decreasing temperature, a zero-bias
anomaly develops in $G(V, T)$, the height of which is
plotted in Fig.~\ref{fig:fig4}. For $T < T_K$, the anomaly
acquires a $\sqrt{V}$ voltage dependence, which is rounded off for
voltages on the scale of the temperature. For $T \ll T_K$,
one is left with a sharp cusp in $G(V, T)$ at zero bias.
Similar to the $\sqrt{T}$ temperature dependence of the
conductance, also the $\sqrt{V}$ voltage dependence of the
differential conductance stems from the anomalous square-root
energy dependence of the imaginary part of the
conduction-electron $T$-matrix in the two-channel Kondo effect.

\begin{figure}[tb]
\centerline{\epsfxsize=75mm \epsfbox {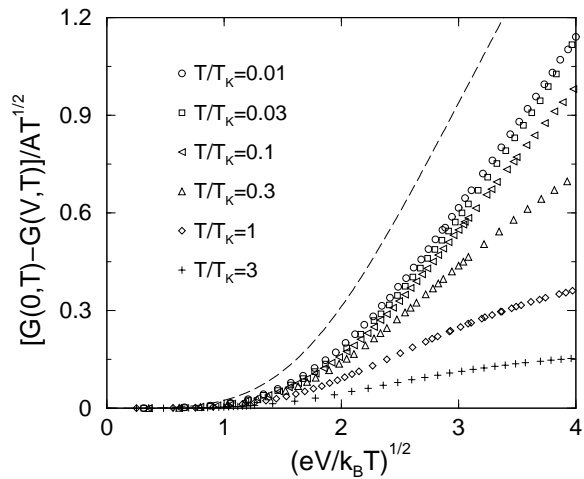}}
\vspace{5pt}
\caption{The function $H = [G(0,T)-G(V,T)]/A T^{1/2}$ versus
	 $(eV/k_BT)^{1/2}$, for the model of Eq.~(\ref{H_2ch}) with
	 symmetric coupling. Here $g_L = g_R = 0.02$, while
	 $A$ is extracted from the leading square-root
	 temperature dependence of the conductance:
	 $G(0, T) = G(0,0) - A \sqrt{T}$. For $T \ll T_K$, all
	 curves collapse onto a single scaling curve.
	 Deviations from scaling are seen for $T$ as low as
	 $0.3 T_K$. At higher temperatures the curves fan out,
	 signaling departure from the scaling regime of the
	 two-channel Kondo effect. The
         dashed line shows for comparison the corresponding
	 scaling curve for a noninteracting tunnel junction
	 with a square-root energy-dependent transmission
	 coefficient (see text).}
\label{fig:fig6}
\end{figure}

Perhaps the most distinctive signature of the two-channel Kondo
effect, though, lies in the scaling form of the differential
conductance with $eV/k_B T$. Following Ralph
{\em et al}.,~\cite{RLvDB94} we plotted in Fig.~\ref{fig:fig6}
the scaling function
\begin{equation}
H(V, T) = \frac{G(0, T) - G(V, T)}{A \sqrt{T}}
\label{2ch_scaling}
\end{equation}
versus $x = (eV/k_B T)^{1/2}$, where $A$ is extracted from
the leading temperature dependence of the conductance:
$G(0, T) = G(0,0) - A \sqrt{T}$ [i.e., $A$ is related to $B$
of Eq.~(\ref{2ch_conductance}) through $A = B/\sqrt{T_K}$].
For $T \ll T_K$, all curves collapse onto a single scaling
curve, indicating that $H(V, T)$ reduces to a function of
the single scaling variable $eV/k_B T$. The resulting scaling
curve features two qualitatively different regimes:
$x \alt 1$ and $x \gg 1$. For $x \alt 1$, the temperature
cuts off the two-channel Kondo effect. Hence $H(x)$ is
proportional to $x^4$, corresponding to a quadratic voltage
dependence of the differential conductance.
By contrast, for $x \gg 1$ the voltage bias serves as the
cut-off energy, and $H(x)$ depends linearly on $x$. The
crossover between these two regimes takes place for $x \sim 1$.
With increasing $T$ there are deviations from scaling. These
are significant for $T$ as low as $0.3 T_K$. At higher
temperatures the curves fan out, signaling departure from the
scaling regime of the two-channel Kondo effect.

Note that the scaling curve of Fig.~\ref{fig:fig6} is quite
similar to the one computed for tunneling through a two-channel
Anderson impurity,~\cite{HKH94} and the one measured for
zero-bias anomalies in ballistic metallic point contacts.~\cite{RLvDB94}
While not unexpected, this resemblance of scaling curves is by
no means obvious, since the voltage bias couples differently
to the impurity within the two-channel Anderson model of Hettler
{\em et al}.~\cite{HKH94} and the Hamiltonian of Eq.~(\ref{H_2ch}).
In Ref.~\onlinecite{HKH94}, the spin-up and spin-down
conduction-electron sectors undergo the same chemical-potential
splitting, as the bias couples to the two spin sectors in an
identical manner. By contrast, only the lead electrons (the
isospin-up sector)
experience a chemical-potential splitting within the Hamiltonian
of Eq.~(\ref{H_2ch}), which breaks the equivalence of the two
isospin sectors. As indicated by Fig.~\ref{fig:fig6}, this
qualitative difference in the coupling to the voltage bias
has only a moderate effect on the scaling curve, which is
more shallow in Fig.~\ref{fig:fig6} than in Ref.~\onlinecite{HKH94}.

That the shape of the scaling curve depends only moderately
on the microscopic details of the system can be understood from
comparison to a simple toy model, consisting of a noninteracting
tunnel junction with a square-root energy-dependent transmission
coefficient: ${\cal T}(\epsilon) = {\cal T}_0 - {\cal T}_1
|\epsilon|^{1/2}$. In this case one can compute the scaling
function of Eq.~(\ref{2ch_scaling}) exactly to
obtain~\cite{comment_on_nonint}
\begin{equation}
H(V, T) = \frac{F(eV/k_B T)}{F(0)} - 1 ,
\label{Toy_model_H}
\end{equation}
with
\begin{equation}
F(y) = \int_{-\infty}^{\infty}
       \frac{e^x}{(1 + e^x)^2} \sqrt{|x -y|}\ dx .
\label{Toy_model_F}
\end{equation}
Note that Eqs.~(\ref{Toy_model_H}) and (\ref{Toy_model_F}) are
independent of both ${\cal T}_0$ and ${\cal T}_1$. As seen in
Fig.~\ref{fig:fig6}, the scaling curve of Eq.~(\ref{Toy_model_H})
has the same general structure as that of the Hamiltonian of
Eq.~(\ref{H_2ch}), despite the fact that the noninteracting model
does not account for the strong temperature and voltage dependence
of the effective transmission coefficient of Eq.~(\ref{T_eff_2ch}).
Thus, while the precise shape of the scaling curve does depend on
the microscopic details of the system at hand, its general structure
appears to be dictated by the square-root energy dependence of
the imaginary part of the conduction-electron $T$-matrix at zero
temperature and zero bias.

\begin{figure}[tb]
\centerline{\epsfxsize=75mm \epsfbox{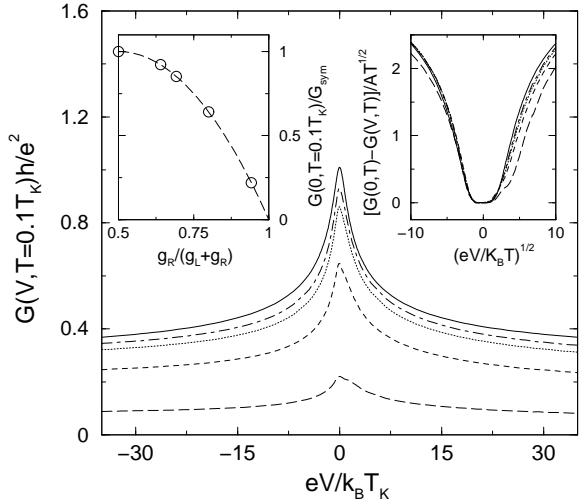}}
\vspace{5pt}
\caption{The differential conductance $G(V, T) = dI/dV$ versus
         voltage bias, for the model of Eq.~(\ref{H_2ch}) with
	 different ratios of the couplings to the right and
	 left leads, $x = g_R/(g_R + g_L)$. Here $g = 0.04$ is
	 kept fixed in all curves as to maintain the same Kondo
	 temperature $k_B T_K/D = 3.85\times 10^{-3}$. The
	 temperature is equal to $T/T_K = 0.1$, while
	 $G_{\rm sym} = 1.01e^2/h$ is the conductance for
	 symmetric coupling at $T/T_K = 0.1$.
	 In going from up to down, $x$ takes the
	 values $x = 0.5, 0.64, 0.69, 0.8$, and $0.94$. With
	 increasing asymmetry, $G(V, T)$ is reduced. Reduction of
	 the zero-bias conductance is according to the standard
	 formula $4x(1-x)$ [left inset: circles are the computed
	 KNCA conductances, the dashed curve marks the parabola
	 $4x(1-x)$]. As a function of bias, an asymmetry develops
	 in the differential conductance. This asymmetry is an
	 artifact of the KNCA. It is emphasized in the right
	 inset, where the scaling function $[G(0,T)-G(V,T)]/AT^{1/2}$,
	 with $A$ extracted separately for each curve, is
	 plotted versus $eV/K_BT$.}
\label{fig:fig7}
\end{figure}

\subsubsection{Asymmetric coupling}

Thus far we have focused on symmetric coupling to the right
and left leads. Below we consider the effect of an asymmetry
in the coupling to the two leads. Specifically, we vary
$x = g_R/( g_L + g_R )$ while keeping $g = g_L + g_R$ fixed,
as to maintain the same Kondo temperature $T_K$. Our results
are summarized in Fig.~\ref{fig:fig7}.

The most notable effect of asymmetric coupling is the
reduction in the differential conductance with increasing
asymmetry. As expected, the zero-bias conductance varies with
$x$ according to $4x(1-x)$, which stems from the fact that,
in equilibrium, $A_{\rm imp}(\epsilon)$ in Eq.~(\ref{T_eff_2ch})
depends solely on the sum $g = g_L + g_R$, and not on the
individual $g_{\alpha}$'s. This
effect is well captured by the KNCA (Fig.~\ref{fig:fig7}, left
inset), which serves yet as another check for our numerical
procedure.

However, the KNCA produces an artificial asymmetry in the
voltage dependence of the differential conductance. To see
this we note that for a symmetric density of states,
$\rho_{\alpha}(\epsilon) = \rho_{\alpha}(-\epsilon)$, the
particle-hole transformation $c \rightarrow c^{\dagger}$,
$(S_x,S_y,S_z) \rightarrow (-S_x,S_y,-S_z)$ converts $V$ to
$-V$ in both the Hamiltonian of Eq.~(\ref{H_2ch}) and the initial
density matrix of Eq.~(\ref{rho_0}).~\cite{comment_on_ph_trans}
Since the current operator of Eq.~(\ref{I_operator}) is
antisymmetric under this transformation, one has the
identity $I_{\alpha}(V) = -I_{\alpha}(-V)$. Hence the
differential conductance is a symmetric function of
the bias, regardless of whether $g_L$ and $g_R$ are
equal or not.

It is easy to verify that the KNCA equations preserve this
symmetry of the differential conductance in the case of
symmetric coupling to the two leads. This stems from the
fact that, within the KNCA, the voltage bias enters the
effective transmission coefficient of Eq.~(\ref{T_eff_2ch})
only through the $\Sigma_{+}$ and $P_{+}$ bubbles of
Figs.~\ref{fig:fig2}(a) and \ref{fig:fig2}(c), respectively.
Since the two leads enter $\Sigma_{+}$ and $P_{+}$ with
equal weights for $g_R = g_L$, and since inverting the
sign of $V$ simply corresponds to interchanging the roles
of the two leads, then the resulting bubbles, and thus
${\cal T}_{\rm eff}(\epsilon)$, remain unchanged upon
flipping the sign of $V$. This is no longer the case for
asymmetric coupling, when the two leads enter each of
$\Sigma_{+}$ and $P_{+}$ with
different weights. Indeed, as seen in Fig.~\ref{fig:fig7},
for $g_R \neq g_L$ the KNCA differential conductance
acquires an asymmetry that increases with increasing $x$.
Note that similar (though more pronounced) asymmetries
in the differential conductance were obtained within
the NCA for tunneling through a two-channel Anderson
impurity,~\cite{HKH94} where $G(V,T) = G(-V, T)$ is not
guaranteed by symmetry.

\begin{figure}[tb]
\centerline{\epsfxsize=65mm \epsfbox {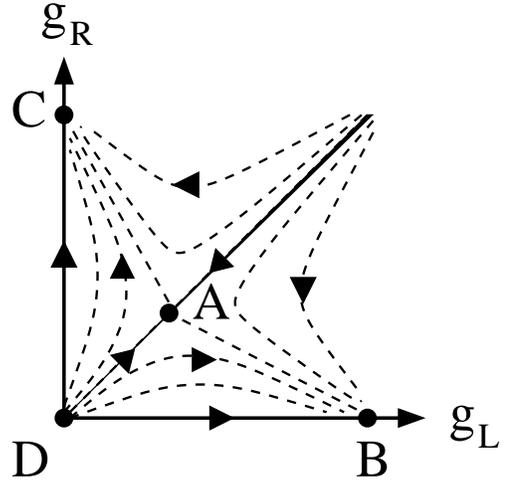}}
\vspace{10pt}
\caption{Schematic renormalization-group (RG) flow diagram for
	 the Hamiltonian of Eq.~(\ref{H_4ch}) in equilibrium.
         Depending on the ratio of $g_L$ to $g_R$, the system
	 flows to one of three fixed points. For symmetric
	 coupling, $g_L = g_R$, the system flows to the
	 four-channel fixed point denoted by A. For any
	 asymmetry in the couplings, the system flows to
	 one of two two-channel fixed points (B or C), at
	 which the initially weaker $g_{\alpha}$ vanishes,
	 while the initially stronger coupling renormalizes to
	 the intermediate-coupling fixed-point value of the
	 two-channel Kondo effect. Thus, the box is strongly
	 coupled to one lead, but effectively decoupled from
	 the other lead. The unstable weak-coupling fixed
	 point is denoted by D.}
\label{fig:fig8}
\end{figure}

\subsection{No coherent propagation between the two junctions}

We proceed with the case where no coherent propagation is allowed
between the leads, i.e., the Hamiltonian of Eq.~(\ref{H_4ch}).
For general couplings, this Hamiltonian reduces in equilibrium
to the planner four-channel Kondo model with channel anisotropy.
Specifically, within this mapping one has $\rho_0 J^{L}_{\perp}
= 2\sqrt{g_L}$ for the two channels associated with the left
junction, and $\rho_0 J^{R}_{\perp} = 2\sqrt{g_R}$ for the two
channels associated with the right junction. Only for $g_L = g_R$
is an isotropic planner four-channel Kondo model recovered,
which shows the qualitative difference between symmetric
and asymmetric couplings. For symmetric coupling, the model
flows in equilibrium to the four-channel Kondo fixed point,
while for any asymmetry it flows to a two-channel
Kondo fixed point where one lead --- that with the weaker
bare $g_{\alpha}$ --- is effectively decoupled. This
renormalization-group (RG) flow diagram of the model is
illustrated in Fig.~\ref{fig:fig8}. As emphasized by Furusaki
and Matveev,~\cite{FM95} the resulting conductance identically
vanishes at $T = 0$ for any $g_L \neq g_R$.

\begin{figure}[tb]
\centerline{\epsfxsize=75mm \epsfbox{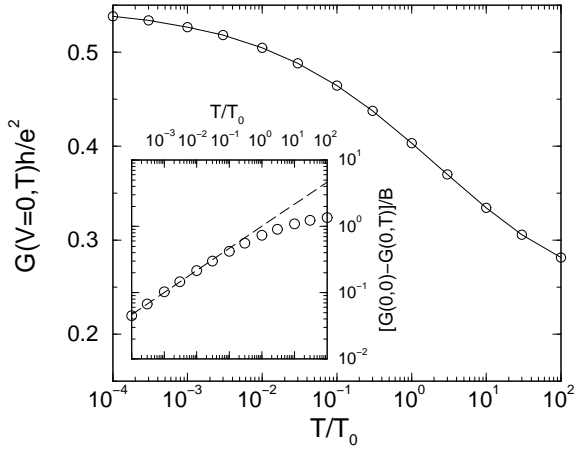}}
\vspace{5pt}
\caption{Temperature dependence of the zero-bias conductance,
         $G(V=0,T)$, for the model of Eq.~(\ref{H_4ch}) with
         $g_L = g_R=0.017$. Here $k_B T_0/D =4.7 \times 10^{-5}$
	 is a characteristic energy scale of order the Kondo
	 temperature. Similar to the two-channel case of
	 Eq.~(\ref{H_2ch}), the conductance monotonically
	 increases with decreasing temperature, crossing over
	 from an approximate logarithmic temperature dependence
	 at intermediate temperatures to a $T^{1/3}$ behavior at
	 low temperatures. The latter behavior is demonstrated
	 in the inset, where a log-log plot of
	 $[G(0, 0) - G(0, T)]/B$ versus $T/T_0$ is shown (open
	 circles) and compared to $(T/T_0)^{1/3}$ (dashed line).
	 Here $G(0,0) = 0.547e^2/h$ and $B = 0.195e^2/h$ were
	 extracted from a fit to the $T^{1/3}$ temperature
	 dependence of Eq.~(\ref{4ch_conductance}).}
\label{fig:fig9}
\end{figure}

From the discussion above it is clear that, unlike in the
case of the Hamiltonian of Eq.~(\ref{H_2ch}), there is no
single energy scale that governs the low-temperature,
low-bias transport properties of the system for all ratios of
$g_L$ to $g_R$. If for symmetric coupling, $g_L = g_R = g$,
transport is governed at low energies by the four-channel
Kondo temperature of Eq.~(\ref{T_K_4ch}),
for asymmetric coupling there is a new crossover scale
for the onset of two-channel behavior. Below we explore
both regimes.

\subsubsection{Symmetric coupling}

Figure~\ref{fig:fig9} shows the temperature dependence of
the zero-bias conductance, for $g_L = g_R = 0.017$. In
the absence of a precise procedure for extracting the
KNCA Kondo temperature in the four-channel case, we use
the scale $k_B T_0 = 4.7 \times 10^{-5}D$ as a rough
estimate of $T_K$. This estimate builds upon the fact
that the KNCA four-channel Kondo temperature is
proportional to $\exp [-2/\rho_0 J_{\perp}]$ rather than
$\exp [-\pi/2\rho_0 J_{\perp}]$,~\cite{LSZ01} which translates
to $T_K^{(KNCA)} \propto \exp [-1/\sqrt{g}]$ in the notation
of Eq.~(\ref{T_K_4ch}). We expect $T_0$ and the actual
KNCA Kondo temperature to be related by a factor of order unity.

\begin{figure}[tb]
\centerline{\epsfxsize=75mm \epsfbox {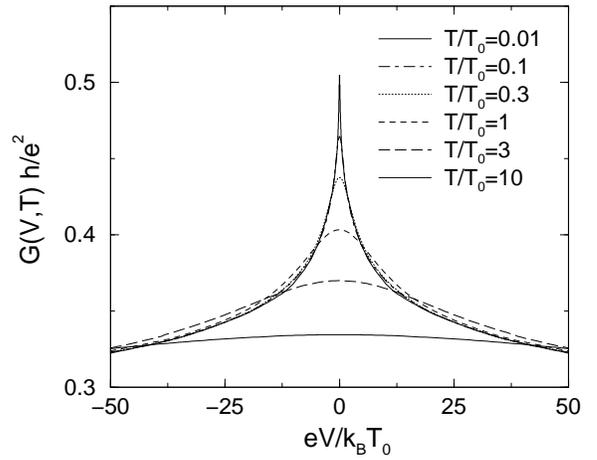}}
\vspace{5pt}
\caption{The differential conductance $G(V, T) = dI/dV$ versus
         voltage bias, for the model of Eq.~(\ref{H_4ch}) with
         $g_L = g_R = 0.017$ and different temperatures. Here
	 $k_B T_0/D$ is equal to $4.7 \times 10^{-5}$. Similar
	 to the case of the Hamiltonian of Eq.~(\ref{H_2ch}), a
	 zero-bias anomaly develops in $G(V, T)$ with decreasing
	 temperature. At low temperature, the anomaly acquires
	 a $V^{1/3}$ voltage dependence, in accordance with the
	 onset of the four-channel Kondo effect.}
\label{fig:fig10}
\end{figure}

In accordance with the onset of the four-channel Kondo
effect, the conductance is enhanced with decreasing temperature,
approaching the zero-temperature value of $G(0, 0) = 0.547e^2/h$.
Somewhat below $T_0$, the conductance crosses over from
an approximate logarithmic temperature dependence to a
$T^{1/3}$ power-law form. The latter behavior is demonstrated
in the inset to Fig.~\ref{fig:fig9}, where the low-$T$
conductance is successfully fitted to
\begin{equation}
G(0, T) = G(0, 0) - B \left(\frac{T}{T_K}\right)^{1/3}
\label{4ch_conductance}
\end{equation}
with $G(0,0) = 0.547e^2/h$ and $B = 0.195e^2/h$.


\begin{figure}[bt]
\centerline{\epsfxsize=75mm \epsfbox {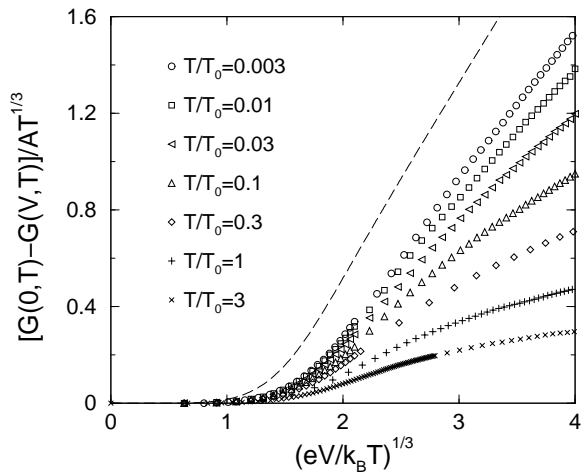}}
\vspace{5pt}
\caption{The function $H = [G(0,T)-G(V,T)]/A T^{1/3}$ versus
	 $(eV/k_BT)^{1/3}$, for the model of Eq.~(\ref{H_4ch})
	 with $g_L = g_R = 0.017$. Here $A$ is extracted from
	 the leading $T^{1/3}$ temperature dependence of the
	 conductance: $G(0, T) = G(0,0) - A T^{1/3}$. For
	 $T \ll T_0$, all curves converge to a single scaling
	 curve. However, pronounced deviations from scaling
	 persist down to much lower temperatures than for
	 the Hamiltonian of Eq.~(\ref{H_2ch}) (compare with
	 Fig.~\ref{fig:fig6}). The dashed line
	 shows for comparison the corresponding scaling curve
	 for a noninteracting tunnel junction with an
	 $|\epsilon|^{1/3}$ energy-dependent transmission
	 coefficient.}
\label{fig:fig11}
\end{figure}

The anomalous $T^{1/3}$ temperature dependence of the conductance
at low $T$ is again a manifestation of the $|\epsilon|^{1/3}$
energy dependence of the imaginary part of the zero-temperature
conduction-electron $T$-matrix in the four-channel Kondo
effect.~\cite{CZ98,AL91} The same power law also characterizes
the voltage dependence of the zero-bias anomaly in the differential
conductance, which is plotted in Fig.~\ref{fig:fig10}.

Similar to the model of Eq.~(\ref{H_2ch}), the differential
conductance for the model of Eq.~(\ref{H_4ch}) also displays
scaling with $V/T$, but with a different power law.
Specifically, Fig.~\ref{fig:fig11} shows a plot of the scaling
function
\begin{equation}
H(V, T) = \frac{G(0, T) - G(V, T)}{A T^{1/3}}
\label{4ch_scaling}
\end{equation}
versus $x = (eV/k_B T)^{1/3}$, where $A$ is extracted from the
leading $T^{1/3}$ temperature dependence of the conductance:
$G(0, T) = G(0,0) - A T^{1/3}$ [i.e., $A$ is related to $B$
of Eq.~(\ref{4ch_conductance}) through $A = B/T_K^{1/3}$].
For $T \ll T_0$, all lines
converge to a single scaling curve, indicating that $H(V, T)$
indeed reduces to a function of the single scaling variable
$eV/k_B T$. However, the approach to scaling with decreasing
$T$, as well as the fanning out of the curves with increasing
$T$, is notably slower than for the Hamiltonian of
Eq.~(\ref{H_2ch}) (compare with Fig.~\ref{fig:fig6}). The
slower approach to scaling is consistent with the fact that
the sub-leading energy dependence of the imaginary part of
the $T = 0$ conduction-electron $T$-matrix, which
breaks scaling, is stronger for the four-channel Kondo
effect~\cite{CZ98,AL91} (of order $|\epsilon|^{2/3}$).
The slower fanning out at higher $T$ stems from the weaker
$T^{1/3}$ temperature dependence of the denominator in
Eq.~(\ref{4ch_scaling}), as compared to $\sqrt{T}$ in
Eq.~(\ref{2ch_scaling}).
Thus, scaling with $V/T$ is a sharper diagnostic for the
two-channel Kondo effect that develops for the Hamiltonian
of Eq.~(\ref{H_2ch}) than it is for the four-channel Kondo effect
that develops for the Hamiltonian of Eq.~(\ref{H_4ch}) with
$g_L = g_R$.

As in the case of the Hamiltonian of Eq.~(\ref{H_2ch}),
the general shape of the scaling curve of Eq.~(\ref{4ch_scaling})
is largely dictated by the $|\epsilon|^{1/3}$ energy
dependence of the imaginary part of the zero-temperature,
zero-bias conduction-electron $T$-matrix, $T_{\alpha\alpha}(\epsilon)
= t_{\alpha}^2 G_{{\rm imp}, \alpha}^{r}(\epsilon)$.
To see this, we plotted for
comparison in Fig.~\ref{fig:fig11} the corresponding scaling
function for a simple toy model, consisting of a noninteracting
tunnel junction with an $|\epsilon|^{1/3}$ energy-dependent
transmission coefficient: ${\cal T}(\epsilon) =
{\cal T}_0 - {\cal T}_1 |\epsilon|^{1/3}$. The exact scaling
curve for this toy model remains given by Eq.~(\ref{Toy_model_H}),
but with a slightly modified $F$ function:~\cite{comment_on_nonint}
\begin{equation}
F(y) = \int_{-\infty}^{\infty}
       \frac{e^x}{(1 + e^x)^2} |x -y|^{1/3} dx .
\label{Toy_model_F_4ch}
\end{equation}
Clearly, the noninteracting model lacks the complicated temperature
and voltage dependences of $G_{{\rm imp}, \alpha}^{r}$ and
$G_{{\rm imp}, \alpha}^{<}$ in Eq.~(\ref{I_4ch_exact}).
Nevertheless, the scaling curve of Eqs.~(\ref{Toy_model_H})
and (\ref{Toy_model_F_4ch}) is quite similar to that of the
Hamiltonian of Eq.~(\ref{H_4ch}) with $g_L = g_R$, as seen
in Fig.~\ref{fig:fig11}. Hence the overall shape of the
scaling curve for the Hamiltonian of Eq.~(\ref{H_4ch}) is
largely dictated by the $|\epsilon|^{1/3}$ energy dependence
of $A_{{\rm imp}, \alpha}(\epsilon)$ at zero bias and zero
temperature.

\begin{figure}[tb]
\centerline{\epsfxsize=75mm \epsfbox {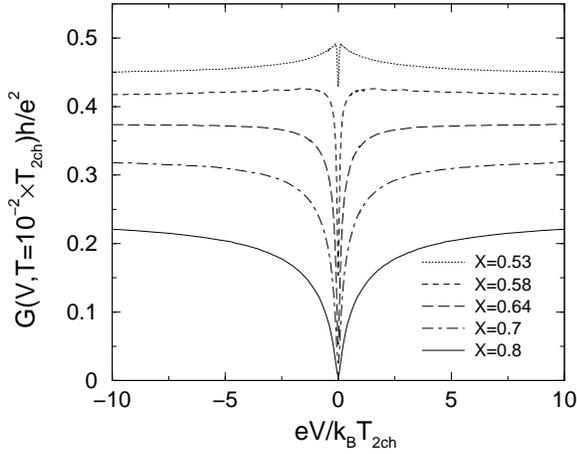}}
\vspace{5pt}
\caption{The differential conductance $G(V, T) = dI/dV$ versus
         voltage bias, for the model of Eq.~(\ref{H_4ch}) with
	 different $x = g_R/(g_R + g_L)$. Here $g_R = 0.04$
	 and $T/T_{\rm 2ch} = 0.01$ are fixed in all curves.
	 As a reference energy, we use the two-channel Kondo
	 temperature obtained for sole coupling to the
	 right lead with $g_R = 0.04$:
	 $k_B T_{\rm 2ch}/D = 3.85\times 10^{-3}$. With
	 increasing $x$, there is a reduction of the overall
	 differential-conductance signal, similar to the one
	 seen in Fig.~\ref{fig:fig7}. A qualitatively different
	 behavior is found near zero bias, where the peak for
	 symmetric coupling ($x = 0.5$) first splits with
	 increasing $x$, leaving only a broadened dip for
	 $x = 0.8$. For $x = 0.53$, the initial enhancement
	 of the differential conductance with decreasing $V$
	 is associated with an initial flow towards the
	 four-channel fixed point (point A in
	 Fig.~\ref{fig:fig8}). The dip at $V = 0$ signals
	 an eventual flow towards the two-channel fixed point
	 where the left is lead decoupled (point C in
	 Fig.~\ref{fig:fig8}). At zero temperature, the
	 conductance vanishes for all $x > 0.5$.}
\label{fig:fig12}
\end{figure}

\begin{figure}[tb]
\centerline{\epsfxsize=75mm \epsfbox{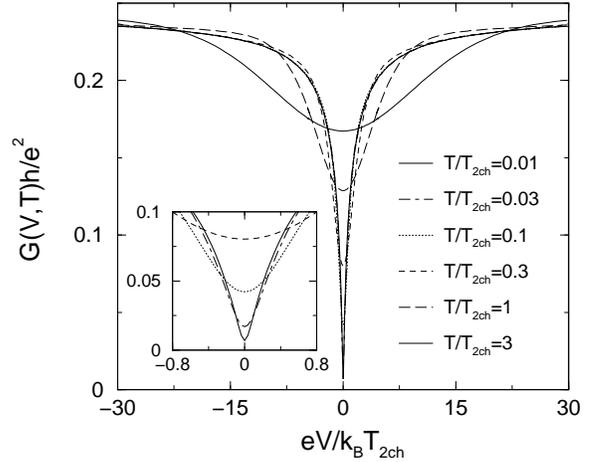}}
\caption{Temperature dependence of the differential conductance
	 $G(V, T) = dI/dV$, for the model of Eq.~(\ref{H_4ch})
	 with $g_R/(g_R + g_L) = 0.8$. Here $g_L = 0.01$ and
	 $g_R = 0.04$. $k_B T_{\rm 2ch}/D = 3.85\times 10^{-3}$
	 is the two-channel Kondo temperature obtained for
	 sole coupling to the right lead with $g_R = 0.04$. With
	 decreasing temperature, a sharp dip develops in the
	 differential conductance at $V = 0$. Specifically,
         $G(0, T)$ is reduced by a factor of $25$ in going
	 from $T/T_{\rm 2ch} = 3$ down to $T/T_{\rm 2ch} = 0.01$,
	 consistent with the flow to the two-channel fixed
	 point where the left lead is decoupled from the box.
	 Inset: an enlarged image of the dip at low bias.}
\label{fig:fig13}
\end{figure}

\subsubsection{Asymmetric coupling}

As discussed above, one expects the zero-temperature conductance
to vanish for any asymmetry in the coupling to the left and right
leads. In the limit of both a large asymmetry and strong tunneling
to one lead (i.e., a nearly open channel), this was indeed shown
to be the case by Furusaki and Matveev.~\cite{FM95} For general
couplings, one expects the system to initially flow towards the
four-channel fixed point denoted by A in Fig.~\ref{fig:fig8},
before curving towards one of the two-channel fixed points,
either B or C in Fig.~\ref{fig:fig8}. Depending on the degree
of asymmetry, this flow may result in a nonmonotonic temperature
and voltage dependence of the differential conductance. Below
we explore this scenario within the KNCA.

Figure~\ref{fig:fig12} shows the evolution of the differential
conductance with increasing $x = g_R/(g_R + g_L)$. Here we have
fixed the coupling to the right lead at $g_R = 0.04$, and varied
$g_L$ from $g_L/g_R = 1$ down to $g_L/g_R = 0.25$. As a reference
energy, we use the two-channel Kondo temperature corresponding
to sole coupling to the right lead, which is the only relevant
low-energy scale for $g_R \gg g_L$. For $g_R = 0.04$ and
$g_L = 0$, this two-channel Kondo temperature is equal to
$k_B T_{\rm 2ch}/D = 3.85\times 10^{-3}$. All curves presented
were computed for $T/T_{\rm 2ch} = 0.01$.
An important point to notice is that the KNCA respects the
symmetry $G(V, T) = G(-V, T)$ for the Hamiltonian of
Eq.~(\ref{H_4ch}) with $g_L \neq g_R$, in contrast to the
Hamiltonian of Eq.~(\ref{H_2ch}). This
is evident from the symmetric differential-conductance
curves of Fig.~\ref{fig:fig12}.

The obvious effect of increasing $x$ in Fig.~\ref{fig:fig12}
is to reduce the overall differential-conductance signal. This
is to be expected, and is also seen in Fig.~\ref{fig:fig7}
for the Hamiltonian of Eq.~(\ref{H_2ch}). A more dramatic
effect takes place near zero bias. Here the peak that forms
for symmetric coupling ($x = 0.5$) first splits with increasing
$x$, leaving only a broadened dip for the larger values of $x$
in Fig.~\ref{fig:fig12}. In particular, for $x$ close to but
larger than $0.5$ (exemplified by $x = 0.53$), a nonmonotonic voltage
dependence of the differential conductance is seen near zero
bias. This behavior is consistent with the notion of an initial
flow towards the four-channel fixed point, before departing
towards the two-channel fixed point with the left lead decoupled.
For larger asymmetries (i.e., larger $x$), the system no longer
flows close to the four-channel fixed point, leaving only a
monotonic dip in $G(V, T)$.

The emergence of a new low-energy scale for asymmetric coupling
is clearly seen in Fig.~\ref{fig:fig12}. For symmetric coupling,
$x = 0.5$, the four-channel Kondo temperature $T_{\rm 4ch}$ is
the only relevant low-energy scale. For small asymmetries, represented
by $x = 0.53$, a new low-energy scale $T_{\rm dip} \ll T_{\rm 4ch}$
emerges, corresponding to half the width of the zero-temperature
dip that opens in $G(V, T)$. The new scale $T_{\rm dip}$ steadily
grows with increasing asymmetry, saturating at
$T_{\rm dip} \sim T_{\rm 2ch}$ for large asymmetries (see, e.g.,
$x = 0.8$ in Figs.~\ref{fig:fig12} and \ref{fig:fig13}).
In this limit, no traces are left of the somewhat smaller
four-channel Kondo temperature $T_{\rm 4ch}$. Due to the small
temperatures involved we are unable to systematically study the
evolution of $T_{\rm dip}$ for small asymmetries. Nevertheless,
by going to lower temperatures we can estimate that $T_{\rm dip}$
for $x = 0.58$ is at least ten-fold smaller than for $x = 0.8$.

\begin{figure}[tb]
\centerline{\epsfxsize=75mm \epsfbox{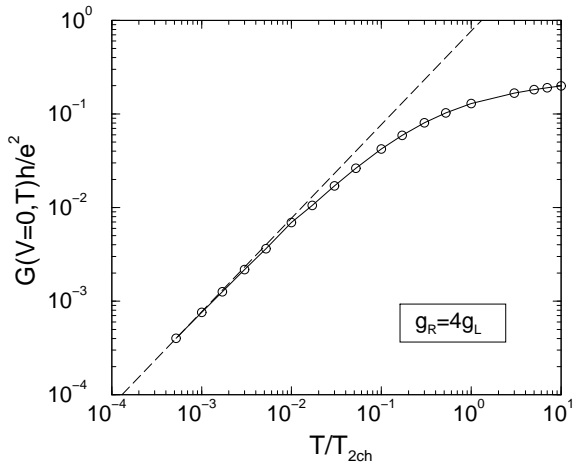}}
\caption{Temperature dependence of the conductance, for the
	 model of Eq.~(\ref{H_4ch}) with $g_L = 0.01$ and
	 $g_R = 0.04$. The conductance monotonically decreases
	 with decreasing temperature, approaching zero for
	 $T \to 0$. Below $T/T_{\rm 2ch} \sim 0.01$, the
	 conductance vanishes linearly with $T$, and is well
	 approximated by
	 $G(0, T) h/e^2 \approx 0.77 T/T_{\rm 2ch}$ (dashed
	 line). Note that the linear temperature dependence
	 only sets in at a temperature much lower than the
	 characteristic width of the dip in the differential
	 conductance (compare with Fig.~\ref{fig:fig13}).}
\label{fig:fig14}
\end{figure}

Figure~\ref{fig:fig12} was obtained for a finite temperature.
At zero temperature, the conductance should vanish for all
$x \neq 0.5$. Although we are unable to directly access the
zero-temperature limit, we can confirm this scenario by
studying the temperature dependence of the differential
conductance. Figures~\ref{fig:fig13} and \ref{fig:fig14}
show our results for $x = 0.8$. For such a large asymmetry,
we are able to go down to temperatures much smaller than
$T_{\rm dip}$. Upon going from $T/T_{\rm 2ch} = 10$ down to
$T/T_{\rm 2ch} = 5\times 10^{-4}$, a $500$-fold reduction
is seen in the conductance. The latter continues to drop
steeply even for $T/T_{\rm 2ch}$ as low as $0.001$, in
accord with a vanishing $T = 0$ conductance. At low
temperatures, $G(0, T)$ vanishes linearly with $T$, in
agreement with previous
predictions.~\cite{FM95,ZZW00} However, the linear-in-$T$
regime is restricted to extremely low temperatures. While
$T_{\rm dip}$ is of the order of $T_{\rm 2ch}$, the linear
temperature dependence of the conductance only sets in at
$T/T_{\rm 2ch} \sim 0.01$, i.e., two orders of magnitude
below $T_{\rm dip}$. Hence the linear-in-$T$ regime is
unlikely to be accessible experimentally, for weak
single-mode junctions.

\section{Discussion}
\label{sec:Discussion}

In this paper, we studied the linear and nonlinear transport
properties of a single-electron transistor at the degeneracy
point. Focusing on weak single-mode junctions, two opposing
scenarios were considered: one by which electrons can propagate
coherently between the two leads, and the other whereby no
electron propagation is allowed between the leads. In the
former scenario, both
leads are assumed to couple to the same mode within the box.
This amounts to a single-mode version of the model used by
K\"onig {\em et al}.~\cite{KSS97} to describe wide tunnel
junctions. In the second scenario, introduced by Furusaki
and Matveev,~\cite{FM95} each lead is coupled to a separate
mode within the box. At the degeneracy point, each of these
models corresponds to a different planner multichannel Kondo
Hamiltonian --- the two-channel Kondo Hamiltonian in the case
where electrons can propagate coherently between the leads,
and the four-channel Kondo Hamiltonian with channel anisotropy
in the case where electron propagation is excluded.

Generalizing the noncrossing approximation to these two
particular nonequilibrium Kondo-type Hamiltonians, distinct
signatures of the multichannel Kondo effect are seen in
the differential conductance. Primarily, for symmetric
coupling, a zero-bias anomaly develops with decreasing
temperature, with a zero-temperature conductance of
order unity. Specifically, within the KNCA we obtain
$G(0,0) = 1.225 e^2/h$ and $G(0,0) = 0.547 e^2/h$,
respectively, for the case where electrons either can or
cannot propagate coherently between the leads. As a
function of temperature, the conductance crosses over
from a characteristic $\ln(T)$ temperature dependence
at intermediate $T$ to a power-law dependence:
$G(0, T) = G(0, 0) - A T^{\eta}$. Here $\eta$ is the
anomalous power law of the leading energy dependence
of the imaginary part of the zero-temperature
conduction-electron $T$-matrix in the appropriate
multichannel Kondo effect: $\eta = 1/2$ for the two-channel
Kondo effect, and $\eta = 1/3$ for the four-channel
Kondo effect.~\cite{CZ98,AL91} The same anomalous power
law also characterizes the low-temperature differential
conductance, which varies as $G(V, T) = G(0, T) - D V^{\eta}$
for $k_B T < e V \ll k_B T_K$.

Scaling of the differential conductance with $eV/k_B T$
is perhaps the clearest fingerprint of the multichannel
Kondo effect that develops.~\cite{HKH94,RLvDB94} As shown
in Figs.~\ref{fig:fig6} and \ref{fig:fig11}, the function
\begin{equation}
H(V, T) = \frac{G(0, T) - G(V, T)}{A T^{\eta}}
\end{equation}
with $k_B T, eV \ll k_B T_K$ reduces to a function of
the single scaling variable $x = (eV/k_B T)^{\eta}$. For
$|x| < 1$, $H(x)$ is proportional to $x^{2/\eta}$,
reflecting the quadratic voltage dependence of $G(V, T)$
for $e V < k_B T$. By contrast, $H(x)$ is linear in $|x|$
for $|x| > 1$, which follows from the $V^{\eta}$ voltage
dependence of $G(V, T)$ for $e V > k_B T$. The crossover
between these two markedly different regimes takes place
for $|x| \sim 1$, i.e., when $e V \sim k_B T$.

As shown in the text, the shape of the resulting scaling curves
can be largely understood within a simple toy model, consisting
of a noninteracting tunnel junction with an $|\epsilon|^{\eta}$
energy-dependent transmission coefficient:
${\cal T}(\epsilon) = {\cal T}_0 - {\cal T}_1 |\epsilon|^{\eta}$.
Although this simplified model lacks the the nontrivial
voltage and temperature dependences of the conduction-electron
$T$-matrix for the two Kondo problems of interest, it does
reproduce the general structure of the two scaling curves.
Hence the latter structures are largely dictated by the
$|\epsilon|^{\eta}$ energy dependence of the imaginary part
of the conduction-electron $T$-matrix at zero temperature and
zero bias. We note, however, that scaling with $eV/k_B T$
is a sharper diagnostic for the two-channel Kondo effect that
develops in the presence of coherent electron propagation, as
compared to the four-channel Kondo effect that develops for
the Hamiltonian of Eq.~(\ref{H_4ch}) with symmetric coupling.

The scaling curve of Fig.~\ref{fig:fig6} is quite similar to
the one obtained for tunneling through a two-channel Anderson
impurity,~\cite{HKH94} where a different coupling of the
impurity to the voltage bias was considered. In the
two-channel Anderson model of Hettler {\em et al.},~\cite{HKH94}
both the spin-up and the spin-down conduction electrons undergo
the same chemical-potential splitting. By contrast, only the
lead electrons (the isospin-up sector) experience a
chemical-potential splitting within the Hamiltonian of
Eq.~(\ref{H_2ch}), which breaks the equivalence of the two
isospin sectors. Despite this qualitative difference
between the two models, the resulting scaling curves
look qualitatively the same, reinforcing the moderate
dependence of the shape of the scaling curve on the
microscopic details of the model.

A clear distinction between the two scenarios under
consideration is revealed when the coupling to the
two leads is asymmetric. For the Hamiltonian of
Eq.~(\ref{H_2ch}), the main effect of an asymmetry
in the coupling is to reduce the zero-temperature
conductance by the conventional factor of
$4 g_L g_R/(g_L + g_R)^2$.
By contrast, for the Hamiltonian of Eq.~(\ref{H_4ch}),
the zero-temperature conductance vanishes for any asymmetry
in the couplings. As first argued by Furusaki and
Matveev,~\cite{FM95} this unexpected result stems from
the fact that the Hamiltonian of Eq.~(\ref{H_4ch}) flows
for any asymmetry to a two-channel Kondo fixed point where
one lead is decoupled from the box. Depending on the degree
of asymmetry, one can obtain then nonmonotonic
differential-conductance curves, as seen in
Fig.~\ref{fig:fig12} for $x = 0.53$.
The nonmonotonicity reflects an initial flow towards the
four-channel fixed point, before curving towards the eventual
two-channel fixed point where one lead is decoupled (see
Fig.~\ref{fig:fig8} for schematic RG trajectories).
For large asymmetries, the system never flows
close to the four-channel fixed point, leaving only a dip
in the low-bias differential conductance. In accordance
with previous predictions,~\cite{FM95,ZZW00} the zero-bias
conductance vanishes linearly with $T$. However, as seen in
Fig.~\ref{fig:fig14}, the linear-in-$T$ regime is restricted
to extremely low temperatures, and is unlikely to be accessible
experimentally for weak single-mode junctions.

Throughout our discussion we have tactfully assumed that the
level spacing inside the box is sufficiently small for a
continuum-limit description to be used. Indeed, there are two
prerequisites for observing a fully developed Kondo effect:
(i) the charging energy $E_C$ must be sufficiently large for
an experimentally accessible Kondo temperature to emerge, and
(ii) the level spacing must be sufficiently small as not to
cut off the multichannel Kondo effect. As emphasized by Zar\'and
{\em et al}.,~\cite{ZZW00} it is difficult to simultaneously
fulfill these two conditions in present-day semiconductor devices
(although some signatures of the two-channel Kondo effect were
recently observed in the charging of a semiconductor quantum
box~\cite{BZAS99}). It is our hope that the host of signatures
provided by this paper will assist in discerning the two
scenarios proposed in the literature, once these experimental
difficulties are overcome.

\section*{Acknowledgments}
We are grateful to Eran Lebanon for many useful discussions
and suggestions. This work was supported in part by the Centers
of Excellence Program of the Israel science foundation, founded
by the Israel Academy of Sciences and Humanities.

\appendix

\section{KNCA equations in the absence of coherent propagation
	 between the leads}
\label{app:KNCA_4ch}

Equations~(\ref{D^r_2ch})--(\ref{S^<_2ch}) were derived for
the Hamiltonian of Eq.~(\ref{H_2ch}). In this appendix, we
specify the corresponding KNCA equations for the Hamiltonian
of Eq.~(\ref{H_4ch}).

The main modification to the KNCA equations for the Hamiltonian
of Eq.~(\ref{H_4ch}) comes from the ladder propagators
$D^{(\lambda)}_{\pm}$, which carry an additional lead index
$\alpha = L, R$. Indeed, contrary to the Hamiltonian of
Eq.~(\ref{H_2ch}), both the spin and lead indices are conserved
along each ladder for the Hamiltonian of Eq.~(\ref{H_4ch}).
In the absence of an applied magnetic field, when the two
spin orientations are equivalent, the resulting ladders are
independent of the spin index, but do depend on the lead
index.

The adaptation of Eqs.~(\ref{D^r_2ch})--(\ref{P^<_2ch_}) to the
Hamiltonian of Eq.~(\ref{H_4ch}) reads
\begin{equation}
D^{r}_{\alpha \pm}(\epsilon) = 
              \frac{ g_{\alpha} P^{r}_{\alpha \mp}(\epsilon)}
              {1 - g_{\alpha} P^r_{\alpha +}(\epsilon)
              P^r_{\alpha -}(\epsilon)} ,
\label{D^r_4ch}
\end{equation}
\begin{equation}
D^{<}_{\alpha \pm}(\epsilon) = 
              \frac{g_{\alpha} P^<_{\alpha \mp}(\epsilon) +
                     g_{\alpha}^2 P^<_{\alpha \pm}(\epsilon)
              \left| P^r_{\alpha \mp}(\epsilon) \right|^2 }
              {\left| 1 - g_{\alpha}
		      P^r_{\alpha +}(\epsilon)
		      P^r_{\alpha -}(\epsilon) \right|^2} \ ,
\end{equation}
where $g_L$ and $g_R$ are the dimensionless tunneling
conductances for the left and right junction, defined in
Eq.~(\ref{g_alpha_4ch}), and
\begin{eqnarray}
P^r_{\alpha +}(\epsilon) &=& -\int_{-\infty}^{\infty}
          G^r_{+}(\epsilon') f(\epsilon'-\epsilon-\mu_{\alpha})
\nonumber \\
&& \;\;\;\;\;\;\;\;\;\;\;
          \times\,
          \nu_{\alpha} (\epsilon'-\epsilon-\mu_{\alpha})
	  d\epsilon' ,
\\
P^<_{\alpha +}(\epsilon) &=& -\int_{-\infty}^{\infty}
          G^<_{+}(\epsilon') f(-\epsilon'+\epsilon+\mu_{\alpha})
\nonumber \\
&& \;\;\;\;\;\;\;\;\;\;\;
          \times\,
          \nu_{\alpha}(\epsilon'-\epsilon-\mu_{\alpha})
	  d\epsilon' ,
\\
P^r_{\alpha -}(\epsilon) &=& -\int_{-\infty}^{\infty}
          G^r_{-}(\epsilon') f(\epsilon'-\epsilon)
          \nu_{B \alpha}(\epsilon'-\epsilon)
	  d\epsilon' ,
\\
P^<_{\alpha -}(\epsilon) &=& -\int_{-\infty}^{\infty}
          G^<_{-}(\epsilon') f(-\epsilon'+\epsilon)
          \nu_{B \alpha}(\epsilon'-\epsilon)
	  d\epsilon' .
\end{eqnarray}
Here we have included the matrix elements of the two end-point
vertices within the $D_{\alpha \pm}$ ladders, and multiplied
$D_{\alpha +}$ and $D_{\alpha -}$ by $\rho_{\alpha}(0)$ and
$\rho_{B \alpha}(0)$, respectively. $\nu_{\alpha}(\epsilon)
= \rho_{\alpha}(\epsilon)/\rho_{\alpha}(0)$ and
$\nu_{B\alpha}(\epsilon) = \rho_{B\alpha}(\epsilon)/\rho_{B\alpha}(0)$
with $\alpha = L, R$ are the reduced density of states in the
leads and in the box. The resulting pseudo-fermion self-energies
acquire the form
\begin{eqnarray}
\Sigma^r_+(\epsilon) &=& -2 \sum_{\alpha = L, R}
	 \int_{-\infty}^{\infty} D^r_{\alpha +}(\epsilon')
	 f(-\epsilon+\epsilon'+\mu_{\alpha}) 
\nonumber \\
&& \;\;\;\;\;\;\;\;\;\;\;\;\;\;\;\;\;\;\;\;\;\;\;
         \times\,
         \nu_{\alpha}(\epsilon-\epsilon'-\mu_{\alpha})
	 d\epsilon' ,
\\
\Sigma^<_+(\epsilon) &=& -2 \sum_{\alpha = L, R}
	 \int_{-\infty}^{\infty} D^<_{\alpha +}(\epsilon')
	 f(\epsilon-\epsilon'-\mu_{\alpha})
\nonumber \\
&& \;\;\;\;\;\;\;\;\;\;\;\;\;\;\;\;\;\;\;\;\;\;\;
         \times\,
         \nu_{\alpha}(\epsilon-\epsilon'-\mu_{\alpha})
	 d\epsilon' ,
\\
\Sigma^r_-(\epsilon) &=& -2 \sum_{\alpha = L, R}
	 \int_{-\infty}^{\infty} D^r_{\alpha -}(\epsilon')
	 f(-\epsilon+\epsilon')
\nonumber \\
&& \;\;\;\;\;\;\;\;\;\;\;\;\;\;\;\;\;\;\;\;\;\;\;
         \times\,
	 \nu_{B \alpha}(\epsilon-\epsilon')
	 d\epsilon' ,
\\
\Sigma^<_-(\epsilon) &=& -2 \sum_{\alpha = L, R}
	 \int_{-\infty}^{\infty} D^<_{\alpha -}(\epsilon')
	 f(\epsilon-\epsilon')
\nonumber \\
&& \;\;\;\;\;\;\;\;\;\;\;\;\;\;\;\;\;\;\;\;\;\;\;
         \times\,
	 \nu_{B \alpha}(\epsilon-\epsilon')
	 d\epsilon' ,
\label{S^<_4ch}
\end{eqnarray}
where the extra factor of two comes from summation
over the two equivalent spin orientations.

Equations~(\ref{D^r_4ch})--(\ref{S^<_4ch}) are the complete
summation of the KNCA class of diagrams for the pseudo-fermion
self-energies, in the case where no coherent propagation
is allowed between the leads [i.e., the Hamiltonian of
Eq.~(\ref{H_4ch})]. They are analogous to
Eqs.~(\ref{D^r_2ch})--(\ref{S^<_2ch}) for the Hamiltonian
of Eq.~(\ref{H_2ch}).


\begin{thebibliography}{99}
\bibitem{comment_on_diff_cond} The nonlinear differential conductance
        of the single-electron transistor was considered, for example,
	in Ref.~\onlinecite{KSS97}, for the case where electrons can
	propagate coherently between the two leads. However, neither
	of the approximations used (low-order perturbation theory
	or the resonant-tunneling approximation) adequately describes
	the Kondo regime for weak single-mode tunnel junctions.
\bibitem{comment_on_Tk} Equation~(\ref{T_K_2ch}) for $T_K$ is obtained
	from Eq.~(63) of Ref.~\onlinecite{Matveev91} by setting $N = 2$
	for the two-channel case, and substituting $\sqrt{g_L + g_R}$
	for $\nu J_0$. The latter correspondence follows from
	the construction of the appropriate ``bonding'' and
	``anti-bonding'' combinations referred to in the main text.
\bibitem{comment_on_4ch_Tk} Equation~(\ref{T_K_4ch}) for $T_K$ is
        obtained from Eq.~(63) of Ref.~\onlinecite{Matveev91} by
	setting $N = 4$ for the four-channel case, and substituting
	$\sqrt{g}$ for $\nu J_0$.
\bibitem{comment_on_nu's} Within the two-channel Kondo representation,
        the requirement that $g \nu_B(\epsilon) = g_L \nu_L(\epsilon)
	+ g_R \nu_R(\epsilon)$ amounts to the conventional
	condition that the two isospin orientations share the same
	conduction-electron density of states. If $g \nu_B(\epsilon)
	\neq g_L \nu_L(\epsilon) + g_R \nu_R(\epsilon)$ then the
	two conduction-electron density of states have different
	functional forms, which has the effect of generating an
	effective magnetic field. This in turn translates to a
	shift in the position of the degeneracy point for the
	Coulomb blockade.
\bibitem{comment_on_mu_ave} The general case where $\mu_L + \mu_R \neq 0$
        is equivalent to setting $\mu_L = -\mu_R = eV/2$ and
	$h \to h - (\mu_L + \mu_R)/2$ [i.e., uniformly shifting
	$\mu_L, \mu_R$, and $h$ by $(\mu_L + \mu_R)/2$].
\bibitem{Comment_on_D} Details of the band cut-offs will affect
        the location of the degeneracy point, as well as the
	value of the Kondo temperature $T_K$.
\bibitem{comment_on_nonint} As in the interacting models of
        Eqs.~(\ref{H_2ch}) and (\ref{H_4ch}), we take the chemical
	potentials in the leads to be $\pm eV/2$.
\bibitem{comment_on_ph_trans} Here we made use of the fact
        that $\mu_L = -\mu_R = eV/2$ and $h = 0$. For nonzero
	$h$, the transformation also converts $h$ to $-h$.
\end{thebibliography}
\end{document}